\documentclass[aip,preprint]{revtex4-1}
\usepackage{physics}
\usepackage{hyperref}
\usepackage{siunitx}
\usepackage{graphicx}
\usepackage{subfigure}
\usepackage{caption}
\begin{document}
\title{The collisional relaxation rate of kappa-distributed plasma with multiple components}
\author{Ran Guo}
\email{rguo@cauc.edu.cn}
\affiliation{\textsuperscript{1)}Department of Physics, College of Science, Civil Aviation University of China, Tianjin 300300, China}

\begin{abstract}
	The kappa-distributed fully ionized plasma with collisional interaction is investigated. The Fokker-Planck equation with Rosenbluth potential is employed to describe such a physical system. The results show that the kappa distribution is not a stationary distribution unless the parameter kappa tends to infinity. The general expressions of collisional relaxation rate of multiple-component plasma with kappa distribution are derived and discussed in specific cases in details. For the purpose of visual illustration, we also give those results numerically in figures. All the results show that the parameter kappa plays a significant role in relaxation rate.
\end{abstract}
\pacs{}
\maketitle

\section{Introduction}
After introduced by Vasyliunas in 1968 \cite{Vasyliunas1968JGR}, kappa distribution, as a non-equilibrium stationary distribution, has been applied to model plenty of space plasmas successfully, for instance, solar corona,\cite{Vocks2008AA,Cranmer2014AJL} solar wind,\cite{Maksimovic1997AA,Leubner2004AJ,Zouganelis2008JGRSP,Stverak2009JGRSP} inner heliosheath,\cite{Livadiotis2011AJa} the planetary magnetosphere,\cite{Mauk2004JGR,Schippers2008JGRSP,Dialynas2009JGRSP} and so on.
Theoretical investigations based on kappa distribution are also interesting topics, which refer to various properties of plasma deviated from Maxwellian equilibrium, such as the discussions of the Debye length,\cite{Gougam2011PP,Livadiotis2014JPP} the definition of temperature,\cite{Livadiotis2009JGRSP,Livadiotis2014E} solitary waves and nonlinear waves in plasmas, \cite{Amour2010PP,Tribeche2012PRE,Das2018AA} the instabilities of plasma,\cite{Liu2009PP,Liu2009PPa,Mace2010JGRSP,Eslami2011PP,Lazar2010MNRAS,Lazar2013MNRAS,Vinas2015JGRSP,Baluku2015PP} transport properties, \cite{Du2013PP,Abbasi2016PP,Wang2017PP,EbneAbbasi2017ASS,Wang2018PP}, and some other works, \cite{Marsch1985PF,Hau2007PP,Gong2012PP,Gong2012PPa} etc.
The kappa velocity distribution is usually written as, \cite{Cranmer1998AJ}
\begin{equation}
	f(\vb{v}) = B_{\kappa} \left[ 1 + \frac{1}{2\kappa-3}\frac{m(\vb{v}-\vb{u})^2}{kT} \right]^{-\kappa-1},
	\label{eq-kappa-pdf}
\end{equation}
where $\vb{v}$ is the velocity vector, $m$ is the mass, $k$ is the Boltzmann constant, $T$ is the temperature, $\vb{u}$ is the overall bulk velocity of plasma, $B_\kappa$ is the normalization factor,
\begin{equation}
	B_{\kappa} = \left[ \frac{m}{(2\kappa-3)\pi k T} \right]^{\frac{3}{2}} \frac{\Gamma(\kappa+1)}{\Gamma \left( \kappa - \frac{1}{2} \right)},
\end{equation}
with the condition $\kappa>3/2$ for the convergence purpose. In the distribution function \eqref{eq-kappa-pdf}, $\kappa$ is a parameter which describes a distance away from equilibrium. And in the limitation $\kappa \rightarrow +\infty$, the distribution function \eqref{eq-kappa-pdf} recovers the Maxwellian distribution. Because of the similarity in mathematical form, it has been considered that the kappa distribution has a very close connection with the $q$-distribution in nonextensive statistical mechanics (NSM). \cite{Leubner2002ASS,Leubner2004PP,Livadiotis2009JGRSP,Livadiotis2015JGRSP} It can be shown that the two distributions are equivalent to each other by the appropriate definition of temperature, \cite{Livadiotis2015JGRSP} which means that some properties of NSM could be utilized in the research of kappa-distributed plasma.

As a widely existed nonthermal stationary state in collisionless space plasma, the kappa distribution can be regarded as a steady-state solution of the collisionless Vlasov equation. In this view, one can derive an equation of $\kappa$ in a nonisothermal physical system with external electromagnetic field,\cite{Du2004PLA,Yu2014AP}
\begin{equation}
	\left( \kappa - \frac{3}{2} \right) k \nabla T = e \left(-\nabla \varphi_c +\frac{\vb{u} \times \vb B}{c} \right),
	\label{eq-dueq}
\end{equation}
where $e$ is the elementary charge, $\varphi_c$ is the Coulomb potential, $c$ is the speed of light and $\mathbf{B}$ is the magnetic induction intensity.
It indicates that the parameter $\kappa$ may be linked to the inhomogeneous temperature as well as the existence of external field.
Besides that, it is also discussed that the kappa distribution can be the stationary solution of many different kinds of Fokker-Planck (FP) equation associated with plasma describing different physical processes, for instance, the FP equation which models a stochastic dynamical process under the generalized fluctuation-dissipation relation,\cite{Du2012JSMTE, Guo2013JSMTE, Guo2014PAa} the FP equation describing wave-particle interaction in a superthermal radiation field,\cite{Hasegawa1985PRL} the FP equation depicting weak turbulence,\cite{Yoon2014JGRSP} the FP equation of accelerated electron in solar flares combining the effects of turbulent acceleration and Coulomb collisions,\cite{Bian2014AJ} etc.
However, there are few works which study on the relation between collision and kappa distribution. Although Marsch and Livi have discussed the relaxation rate of single-component plasma with kappa distribution \cite{Marsch1985PF}, there are still some questions need to be studied. Does the kappa distribution would be broken due to the collision between particles in a general multiple-component plasma?
If it is true what direction does the kappa distribution evolve to? Is that the Maxwellian equilibrium? How fast is that relaxation process?
All the above questions are still unclear in the present theory of kappa distribution.

Therefore, the present work focuses on the influence of the collision in fully ionized plasma consisted by multiple kappa-distributed components. The paper is organized as follows. In section 2, we introduce the FP equation with Rosenbluth potential in order to study the collision effects on kappa distribution. In section 3, we examine whether the kappa distribution is the stationary solution of the FP equation and give the relaxation rate by both analytic expressions and numerical results in figures. In section 4, we make the conclusions and the discussions. At last, some detailed calculations are written in Appendix.

\section{Kinetic equation}
We consider a fully ionized plasma consisted by multiple components in which the interaction between particles is Coulomb collision. In such a kind of plasma, small angle Coulomb scatterings are more probably occurred than a large angle one, while a large deflection is consisted by a series of small angle scattering. So the collisional effects of plasma can be described by the FP equation with Rosenbluth potential in which the collision term is given by, \cite{Krall1973,Gurnett2005}
\begin{equation}
	\left (\pdv[]{f_\alpha}{t} \right)_c = \sum_{\beta} \Gamma_{\alpha \beta} \left[ -\pdv[]{}{\vb{v}} \cdot \left( f_\alpha \pdv[]{h_{\alpha \beta}}{\vb{v}} \right) + \frac{1}{2} \pdv[]{}{\vb{v}} \pdv[]{}{\vb{v}} : \left( f_\alpha \pdv[]{g_{\alpha \beta}}{\vb{v}}{\vb{v}} \right)\right],
	\label{eq-fpe}
\end{equation}
where $f_\alpha$ is the velocity distribution function of the $\alpha$-species particle, the summation of $\beta$ on the right side is over all kinds of particles in plasma, $\Gamma_{\alpha \beta}$ is a factor which reads
\begin{equation}
	\Gamma_{\alpha \beta} = \frac{4\pi n_\beta q_\alpha^2 q_\beta^2}{m_\alpha^2} \ln \Lambda
\end{equation}
with the number density $n_\beta$ and the charge $q_\beta$ of the $\beta$ constituent, the charge $q_\alpha$ and the mass $m_\alpha$ of the $\alpha$ one, and the scattering factor $\ln \Lambda$. The Rosenbluth potential $h_{\alpha \beta}$ and $g_{\alpha \beta}$ is defined as \cite{Krall1973,Gurnett2005}
\begin{equation}
	h_{\alpha \beta} (\vb{v}) = \left( 1+\frac{m_\alpha}{m_\beta} \right) \int \frac{f_\beta (\vb{v'})}{\left| \vb{v} - \vb{v'} \right| } \dd \vb{v'},
	\label{eq-rp-h}
\end{equation}
\begin{equation}
	g_{\alpha \beta} (\vb{v}) = \int f_\beta (\vb{v'}) \left| \vb{v} - \vb{v'} \right| \dd \vb{v'},
	\label{eq-rp-g}
\end{equation}
with the distribution function $f_\beta (\vb{v'})$ for $\beta$-species particle. The Eq.\eqref{eq-fpe} can also be expressed in a general form of FP equation as
\begin{equation}
	\left( \pdv{f_\alpha}{t} \right)_c = - \pdv[]{}{\vb{v}} \cdot \vb{S}_{\alpha},
	\label{eq-fpe-s}
\end{equation}
where $\vb{S}_{\alpha}$ is the current of probability,
\begin{equation}
	\vb{S}_{\alpha} = \left[ f_\alpha \vb{H}_{\alpha}(\vb v) - \frac{1}{2} \pdv[]{}{\vb{v}} \cdot \left( f_\alpha \overleftrightarrow{\vb{G}}_{\alpha}(\vb{v}) \right) \right],
	\label{eq-s-hg}
\end{equation}
where $\vb{H}_{\alpha}(\vb{v})$ is the dynamical friction vector and $\overleftrightarrow{\vb{G}}_{\alpha}(\vb{v})$ is the dynamical diffusion tensor,
\begin{equation}
	\vb{H}_{\alpha}(\vb{v}) \equiv \sum_\beta \Gamma_{\alpha \beta} \pdv[]{h_{\alpha \beta}}{\vb{v}},
\end{equation}
and
\begin{equation}
	\overleftrightarrow{\vb{G}}_{\alpha}(\vb{v}) \equiv \sum_\beta \Gamma_{\alpha \beta} \pdv[]{g_{\alpha \beta}}{\vb{v}}{\vb{v}}.
\end{equation}
For a stationary state $f_\alpha (\vb v)$, the FP collision term $(\partial f_\alpha/\partial t)_c$ should equal zero for all possible velocity $\vb v$; while for non-stationary it represents the relaxation rate of initial distribution $f_\alpha (\vb v)$ at initial time $t=0$. 

\section{Examination of stationary distribution and the relaxation rate}
In order to test whether the kappa distribution is the stationary solution of FP equation, we need to substitute Eq.\eqref{eq-kappa-pdf} into Eq.\eqref{eq-fpe}. Generally speaking, one can assume that each kind of particles in plasma follows the kappa distribution with different temperatures and $\kappa$-indexes,
\begin{equation}
	f_\beta(\vb{v}) = B_{\kappa,\beta} \left( 1 + \frac{1}{2\kappa_\beta-3} \frac{m_\beta v^2}{kT_\beta} \right)^{-\kappa_\beta-1},
	\label{eq-kappa-pdfa}
\end{equation}
where the subscript $\beta$ stands for the physical quantities of $\beta$ component in plasma. And the bulk velocity $\vb{u}$ is omitted in the calculations, because on the right side of the collision term \eqref{eq-fpe} the partial derivative is only with respect to $\vb{v}$, or in other words, $\partial \vb{u} / \partial \vb{v} = 0$.
Based on these simplifications, we firstly calculate the Rosenbluth potential. The function $h_{\alpha \beta} (\vb{v})$ has been obtained,\cite{Abbasi2016PP, Du2018CPP}
\begin{equation}
	h_{\alpha \beta}(\vb{v}) = \left( 1+\frac{m_\alpha}{m_\beta} \right) 4\pi B_{\kappa, \beta} \frac{\kappa_\beta-\frac{3}{2}}{\kappa_\beta} \frac{kT_\beta}{m_\beta} {_2 F_1}\left(\frac{1}{2},\kappa_\beta;\frac{3}{2},-A_{\kappa, \beta}v^2 \right),
	\label{eq-h-kappa}
\end{equation}
where $A_{\kappa,\beta} \equiv m_\beta/[(2\kappa_\beta-3)kT_\beta]$ is an abbreviation, and ${_2 F_1}(a,b;c,z)$ is hypergeometric function defined by,\cite{Olver2010NIST}
\begin{equation}
	{_2 F_1}(a,b;c,z) = \frac{\Gamma(c)}{\Gamma(a) \Gamma(c-a)} \int_0^1x^{a-1} (1-x)^{c-a-1} (1-zx)^{-b} \dd x.
	\label{eq-2f1}
\end{equation}
The function $g_{\alpha \beta}(\vb{v})$ can also be derived by substituting the kappa distribution \eqref{eq-kappa-pdfa} into the integral \eqref{eq-rp-g} (details in Appendix \ref{sec-app-g_ab}),
\begin{equation}
	\begin{split}
		g_{\alpha \beta}(\vb{v}) =& \frac{\pi B_{\kappa,\beta}}{\kappa_\beta A_{\kappa,\beta}}
		\left[
		\frac{2}{3} v^2 {_2 F_1}\left(\frac{3}{2},\kappa_\beta;\frac{5}{2},-A_{\kappa,\beta} v^2 \right)
		+ 2 v^2 {_2 F_1}\left(\frac{1}{2},\kappa_\beta;\frac{3}{2},-A_{\kappa,\beta}v^2 \right) \right. \\
		& \left. + \frac{2}{(\kappa_\beta-1) A_{\kappa,\beta}} (1+A_{\kappa,\beta} v^2)^{-\kappa_\beta+1} \right].
	\end{split}
	\label{eq-g-kappa}
\end{equation}
And then the dynamical friction vector $\vb{H}_{\alpha}(\vb{v})$ and the dynamical diffusion tensor $\overleftrightarrow{\vb{G}}_{\alpha}(\vb{v})$ are obtained by taking partial derivative of $h_{\alpha \beta} (\vb{v})$ and $g_{\alpha \beta}(\vb{v})$,
\begin{equation}
	\vb{H}_{\alpha}(\vb{v}) = \sum_\beta -\frac{4\pi}{3} \Gamma_{\alpha \beta} B_{\kappa,\beta} \left( 1+\frac{m_\alpha}{m_\beta} \right) {_2 F_1}\left(\frac{3}{2},\kappa_\beta+1;\frac{5}{2},-A_{\kappa,\beta}v^2 \right) \vb{v},
	\label{eq-Hi}
\end{equation}
and
\begin{equation}
	\begin{split}
		\overleftrightarrow{\vb{G}}_{\alpha}(\vb{v})=& \sum_\beta 4\pi \Gamma_{\alpha \beta} B_{\kappa,\beta} \left\{
		\overleftrightarrow{\vb{U}} \frac{1}{\kappa_\beta A_{\kappa,\beta}} \left[ \frac{1}{2} {_2 F_1}\left(\frac{1}{2},\kappa_\beta;\frac{3}{2},-A_{\kappa,\beta}v^2 \right) -\frac{1}{6}{_2 F_1}\left(\frac{3}{2},\kappa_\beta;\frac{5}{2},-A_{\kappa,\beta}v^2 \right)\right] \right. \\
		&\left. + \vb{v} \vb{v} \left[ \frac{1}{5} {_2 F_1}\left(\frac{5}{2},\kappa_\beta+1;\frac{7}{2},-A_{\kappa,\beta}v^2 \right) - \frac{1}{3} {_2 F_1} \left(\frac{3}{2},\kappa_\beta+1;\frac{7}{2},-A_{\kappa,\beta}v^2 \right) \right] \right\},
	\end{split}
	\label{eq-Gik}
\end{equation}
where $\overleftrightarrow{\vb{U}}$ is the unit dyadic. With the help of Eqs.\eqref{eq-Hi} and \eqref{eq-Gik}, the probability current $\vb{S_\alpha}$ is derived,
\begin{equation}
	\begin{split}
		\vb{S_{\alpha}}=& \sum_\beta -\frac{4\pi}{3}\Gamma_{\alpha \beta} B_{\kappa,\beta} f_\alpha \vb{v} \frac{m_\alpha}{m_\beta} \left[ {_2 F_1}\left(\frac{3}{2},\kappa_\beta+1;\frac{5}{2},-A_{\kappa,\beta}v^2 \right) \right. \\
			&\left. - \frac{\kappa_\alpha+1}{\kappa_\beta} \frac{\kappa_\beta-\frac{3}{2}}{\kappa_\alpha-\frac{3}{2}} \frac{T_\beta}{T_\alpha} (1+A_{\kappa,\alpha}v^2)^{-1} {_2 F_1}\left(\frac{3}{2},\kappa_\beta;\frac{5}{2},-A_{\kappa,\beta}v^2 \right)  \right].
	\end{split}
	\label{eq-pc}
\end{equation}
Substituting Eq.\eqref{eq-pc} into the FP equation \eqref{eq-fpe}, we derive the FP collision term,
\begin{equation}
	\begin{split}
		\left( \pdv[]{f_\alpha}{t} \right)_c =& \sum_\beta \frac{4\pi}{3} \Gamma_{\alpha \beta} B_{\kappa,\beta} f_\alpha \frac{m_\alpha}{m_\beta} (1+A_{\kappa,\alpha} v^2)^{-1}\left\{ 3(1 + A_{\kappa,\beta}v^2)^{-\kappa_\beta} \left[ \frac{1+A_{\kappa,\alpha}v^2}{1+A_{\kappa,\beta}v^2} -\frac{\kappa_\alpha+1}{\kappa_\beta} \frac{\kappa_\beta - \frac{3}{2}}{\kappa_\alpha - \frac{3}{2}} \frac{T_\beta}{T_\alpha} \right] \right. \\
		&-2 (\kappa_\alpha+1) A_{\kappa,\alpha} v^2 \left[ {_2 F_1}\left(\frac{3}{2},\kappa_\beta+1;\frac{5}{2},-A_{\kappa,\beta}v^2 \right) - \frac{\kappa_\alpha+2}{\kappa_\beta} \frac{\kappa_\beta-\frac{3}{2}}{\kappa_\alpha-\frac{3}{2}} \frac{T_\beta}{T_\alpha} (1+A_{\kappa,\alpha}v^2)^{-1}\right. \\
		&\left. \left. \times {_2 F_1}\left(\frac{3}{2},\kappa_\beta;\frac{5}{2},-A_{\kappa,\beta}v^2 \right)\right] \right\}.
	\end{split}
	\label{eq-fpc}
\end{equation}
The distribution function $f_\alpha$ denotes a stationary state if and only if the partial derivative vanishes, namely $(\partial f_\alpha / \partial t)_c = 0$ for arbitrary $\vb v$. For such a complicated equation \eqref{eq-fpc}, it is difficult to find whether there are some special values of $\kappa$ leading to the zero FP collision term. So several cases would be investigated in the next subsections.

However, if the FP collision term $(\partial f_\alpha/\partial t)_c \neq 0$, which means $f_\alpha$ is not a stationary state and will relax to the equilibrium, then we can study the rate of the relaxation. For the purpose of characterization of the relaxation rate, we adopt the absolute rate of change $(\partial f_\alpha/\partial t)_c$ as well as the relative rate of change,
\begin{equation}
	\frac{1}{f_\alpha} \left( \pdv[]{f_\alpha}{t} \right)_c,
\end{equation}
which describes the relative change of the distribution function in unit time and can be also regarded as the inverse of relaxation time. 

\subsection{Plasma with one component}
In the first place we consider a single-component plasma such as a pure electron plasma. In this case, the collision term \eqref{eq-fpc} can be simplified as
\begin{equation}
	\begin{split}
		\left( \pdv[]{f_\alpha}{t} \right)_c = \frac{4\pi}{3} \Gamma_{\alpha \alpha} B_{\kappa,\alpha} f_\alpha (1+A_{\kappa,\alpha}v^2)^{-1} \left\{ -\frac{3}{\kappa_\alpha}(1+A_{\kappa,\alpha}v^2)^{-\kappa_\alpha} -2(\kappa_\alpha +1) A_{\kappa,\alpha} v^2 \right. \\
		\times \left. \left[ {_2 F_1} \left(\frac{3}{2},\kappa_\alpha +1;\frac{5}{2},-A_{\kappa,\alpha}v^2 \right) - \frac{\kappa_\alpha+2}{\kappa_\alpha}(1+A_{\kappa,\alpha}v^2)^{-1} {_2 F_1} \left(\frac{3}{2},\kappa_\alpha;\frac{5}{2},-A_{\kappa,\alpha}v^2 \right) \right]\right\},
	\end{split}
	\label{eq-ft-1c}
\end{equation}
while the probability current is given by,
\begin{equation}
	\vb S_\alpha = - \frac{4\pi}{3} \Gamma_{\alpha \alpha } B_{\kappa,\alpha} f_\alpha \vb{v} \left[ {_2 F_1} \left( \frac{3}{2},\kappa_\alpha+1;\frac{5}{2},-A_{\kappa,\alpha}v^2 \right) - \frac{\kappa_\alpha+1}{\kappa_\alpha} (1+A_{\kappa,\alpha}v^2)^{-1} {_2 F_1}\left( \frac{3}{2},\kappa_\alpha;\frac{5}{2},-A_{\kappa,\alpha}v^2 \right) \right].
	\label{eq-s-1c}
\end{equation}
For convenience, the FP collision term and the probability current can be rewritten respectively as,
\begin{equation}
	\begin{split}
		\left( \pdv[]{f_\alpha}{t} \right)_c =& \frac{4\pi}{3} \Gamma_{\alpha \alpha} {f_\alpha}^2 \left\{ -\frac{3}{\kappa_\alpha} -2(\kappa_\alpha +1) A_{\kappa,\alpha} v^2 \left[ {_2 F_1} \left(1,\frac{3}{2}-\kappa_\alpha;\frac{5}{2},-A_{\kappa,\alpha}v^2 \right) \right. \right. \\
			&\left.\left. - \frac{\kappa_\alpha+2}{\kappa_\alpha} {_2 F_1} \left(1,\frac{5}{2}-\kappa_\alpha;\frac{5}{2},-A_{\kappa,\alpha}v^2 \right) \right]\right\},
	\end{split}
	\label{eq-ft-1ca}
\end{equation}
\begin{equation}
	\vb S_\alpha = - \frac{4\pi}{3} \Gamma_{\alpha \alpha }  {f_\alpha}^2 (1+A_{\kappa,\alpha}v^2) \vb{v} \left[ {_2 F_1} \left( 1,\frac{3}{2}- \kappa_\alpha;\frac{5}{2},-A_{\kappa,\alpha}v^2 \right) - \frac{\kappa_\alpha+1}{\kappa_\alpha} {_2 F_1}\left( 1,\frac{5}{2}-\kappa_\alpha;\frac{5}{2},-A_{\kappa,\alpha}v^2 \right) \right],
	\label{eq-s-1ca}
\end{equation}
where the relation \cite{Olver2010NIST} of hypergeometric function has been used,
\begin{equation}
	{_2 F_1}(a,b;c,z) = (1-z)^{c-a-b} {_2 F_1}(c-a,c-b;c,z).
\end{equation}

The FP collision term will be analyzed in three situations at $\kappa_\alpha=+\infty$, $3/2$ and other values, because the parameter $\kappa_\alpha \in (3/2,+\infty)$.

\subsubsection{The limit $\kappa_\alpha \rightarrow +\infty$}
For $\kappa_\alpha \rightarrow +\infty$, after taking the limitations
\begin{equation}
	\lim_{\kappa_\alpha \rightarrow +\infty} B_{\kappa,\alpha} f_\alpha = \left( \frac{m_\alpha}{2\pi k T_\alpha} \right)^{3} \exp \left( \frac{m_\alpha v^2}{2k T_\alpha} \right),
\end{equation}
\begin{equation}
	\lim_{\kappa_\alpha \rightarrow +\infty} (1+A_{\kappa,\alpha}v^2)^{-1} = 1,
\end{equation}
\begin{equation}
	\lim_{\kappa_\alpha \rightarrow +\infty} \frac{3}{\kappa_\alpha}(1+A_{\kappa,\alpha}v^2)^{-\kappa_\alpha} = 0,
\end{equation}
\begin{equation}
	\lim_{\kappa_\alpha \rightarrow +\infty} \left[ {_2 F_1}\left( \frac{3}{2},\kappa_\alpha+1;\frac{5}{2},-A_{\kappa,\alpha}v^2 \right) - \frac{\kappa_\alpha+2}{\kappa_\alpha} {_2 F_1}\left( \frac{3}{2},\kappa_\alpha;\frac{5}{2},-A_{\kappa,\alpha}v^2 \right)\right] = 0,
\end{equation}
one derives
\begin{equation}
	\lim_{\kappa_\alpha \rightarrow +\infty} \left( \pdv[]{f_\alpha}{t}\right)_c = 0,
	\label{eq-ft-1cinf}
\end{equation}
and in the same way
\begin{equation}
	\lim_{\kappa_\alpha \rightarrow +\infty} \vb S_\alpha = 0.
\end{equation}
This is obvious because when $\kappa \rightarrow +\infty$ the kappa distribution recovers the Maxwellian one, which is the equilibrium state.

\subsubsection{The limit $\kappa_\alpha \rightarrow \frac{3}{2}$}
When $\kappa_\alpha \rightarrow 3/2$, similarly we need to calculate some limitations. The kappa distribution $f_\alpha$ and some hypergeometric functions in the above equations under the limit $\kappa_\alpha \rightarrow 3/2$ are work out directly,
\begin{equation}
	\lim_{\kappa_\alpha \rightarrow \frac 32}  f_\alpha(\vb v)
	\begin{cases}
		= 0,      & \text{if $v \neq 0$} \\
		= \infty, & \text{if $v=0$}
	\end{cases}
	\label{eq-ft=3/2}
\end{equation}
\begin{equation}
	\lim_{\kappa_\alpha \rightarrow \frac 32} {_2 F_1} \left( 1,\frac{3}{2}-\kappa_\alpha;\frac{5}{2},-A_{\kappa,\alpha}v^2 \right) = 1,
\end{equation}
\begin{equation}
	\lim_{\kappa_\alpha \rightarrow \frac 32} {_2 F_1} \left( 1,\frac{5}{2}-\kappa_\alpha;\frac{5}{2},-A_{\kappa,\alpha}v^2 \right) = 0,
\end{equation}
where another relation \cite{Olver2010NIST} has been used for evaluating the limit of hypergeometric function,
\begin{equation}
	\begin{split}
		\frac{\sin (\pi(b-a))}{\pi \Gamma(c)} &{_2 F_1}(a,b;c,z) = \\
		& \frac{(-z)^{-a}{_2 F_1}(a,a-c+1;a-b+1,z^{-1})}{\Gamma(b) \Gamma(c-a) \Gamma(a-b+1)}  - \frac{(-z)^{-b} {_2 F_1}(b,b-c+1;b-a+1,z^{-1})}{\Gamma(a) \Gamma(c-b) \Gamma(b-a+1)}.
	\end{split}
\end{equation}
After taking these results back into Eq.\eqref{eq-ft-1c}, we obtain
\begin{equation}
	\lim_{\kappa_\alpha \rightarrow \frac{3}{2}} \left( \pdv[]{f_\alpha}{t} \right)_c =
	\begin{cases}
		0, \text{ if $v \neq 0$} \\
		\infty, \text{if $v = 0$}
	\end{cases}
	\label{eq-ft-1c32}
\end{equation}
and the relative relaxation rate,
\begin{equation}
	\lim_{\kappa_\alpha \rightarrow \frac{3}{2}} \frac{1}{f_\alpha} \left( \pdv[]{f_\alpha}{t} \right)_c = - \Gamma_{\alpha \alpha} \frac{5}{v^3} .
	\label{eq-rft-1c32}
\end{equation}

Therefore, in this case, the kappa distribution cannot be regarded as a stationary state.

\subsubsection{Other values of $\kappa_\alpha$}
\begin{figure}[ht]
	\centering
	\subfigure[The negative part of absolute relaxation rate for $\kappa_e=1.6$, $2$, and $5$ respectively.]{
		\includegraphics[width=0.6\textwidth]{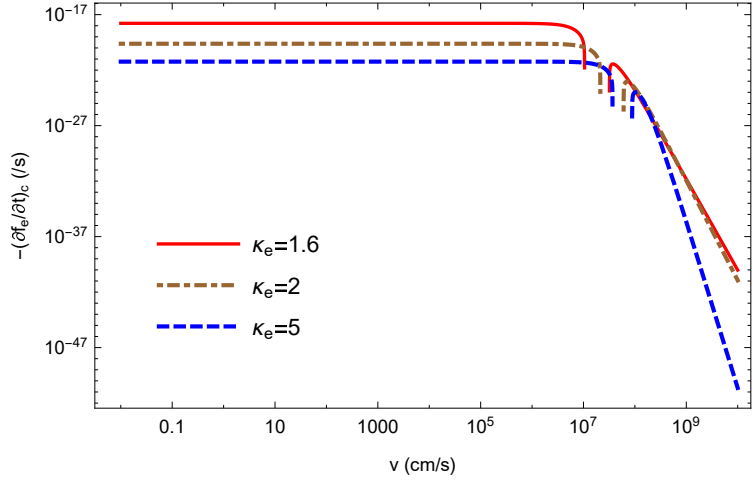}
		\label{ft-1c:a}
	}
	\subfigure[The positive part of absolute relaxation rate for $\kappa_e=1.6$, $2$, and $5$ respectively.]{
		\includegraphics[width=0.6\textwidth]{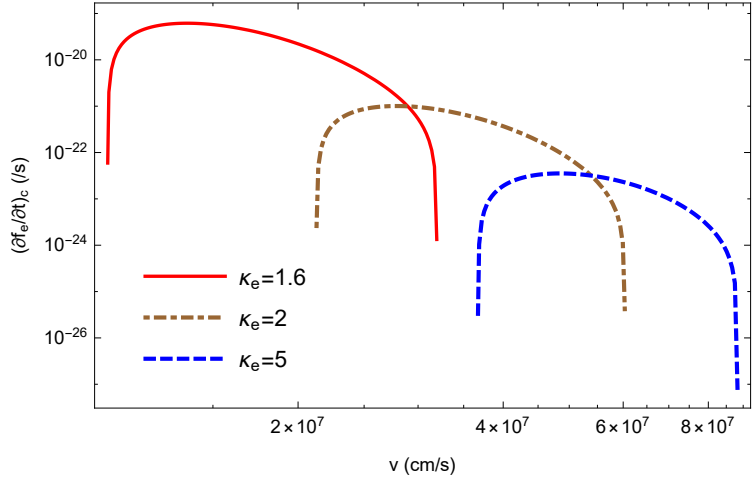}
		\label{ft-1c:b}
	}
	\caption{The absolute relaxation rate $(\partial f_e/\partial t)_c$ of pure electron plasma.}
	\label{ft-1c}
\end{figure}

\begin{figure}[ht]
	\centering
	\subfigure[The negative part of relative relaxation rate for $\kappa_e=1.6$, $2$, and $5$ respectively.]{
		\includegraphics[width=0.6\textwidth]{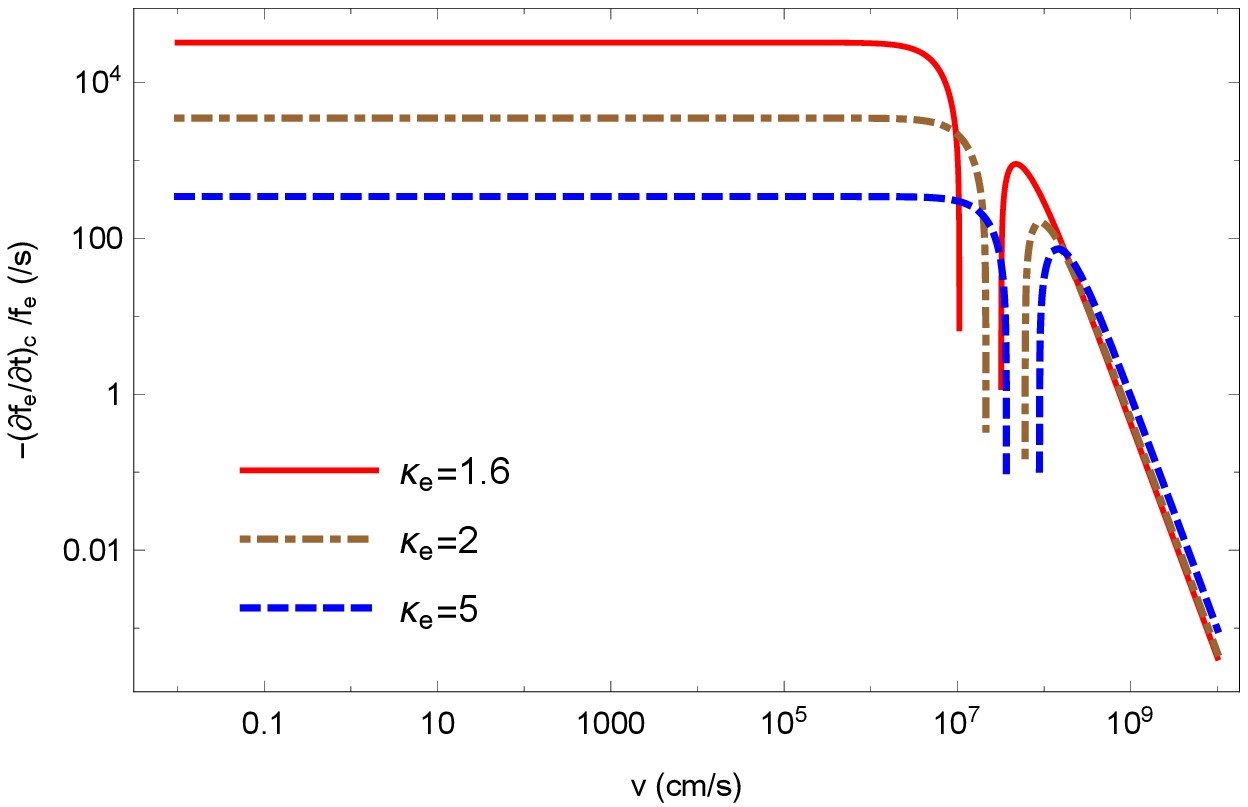}
		\label{rft-1c:a}
	}
	\subfigure[The positive part of relative relaxation rate for $\kappa_e=1.6$, $2$, and $5$ respectively.]{
		\includegraphics[width=0.6\textwidth]{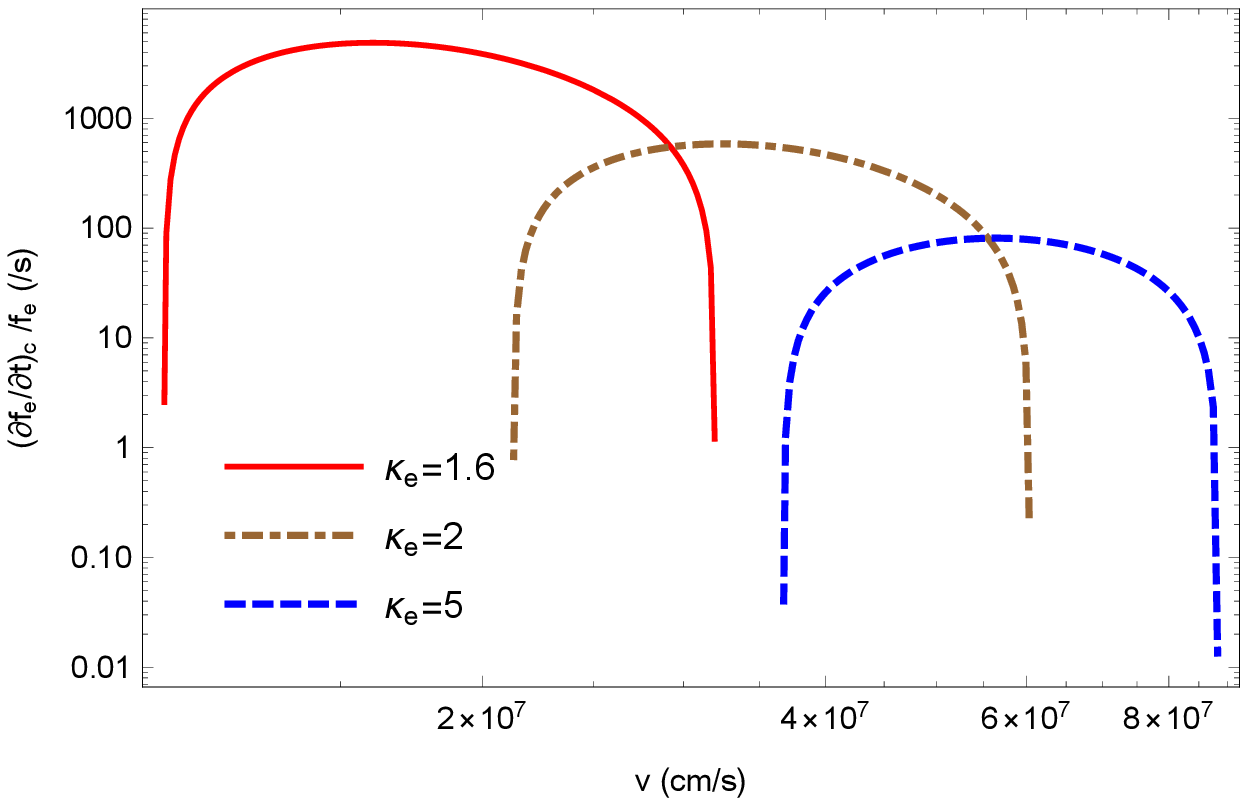}
		\label{rfts-1c:b}
	}
	\caption{The relative relaxation rate $(\partial f_e/\partial t)_c/f_e$ of pure electron plasma}
	\label{rft-1c}
\end{figure}

For other values of $\kappa_\alpha$, the FP collision term \eqref{eq-ft-1c} is a function of $\vb v$, so it cannot be zero for arbitrary $\vb v$ when $\kappa_\alpha \neq \infty$, which means that the kappa distribution is not the stationary solution of the FP equation for single-component plasma. We here take the collisional pure electron plasma as an example in order to show the FP collision term \eqref{eq-ft-1c} for different values of $\kappa_e$ in Figs.\ref{ft-1c} and \ref{rft-1c}, where the typical temperature and density of number in pure electron plasma are set as, $T_e= 10^4 \si{K}$, $n_e=10^8 \si{cm^{-3}}$. And additionally we let the scatter factor $\ln \Lambda=1$.

\begin{figure}[ht]
	\centering
	\includegraphics[width=0.6\textwidth]{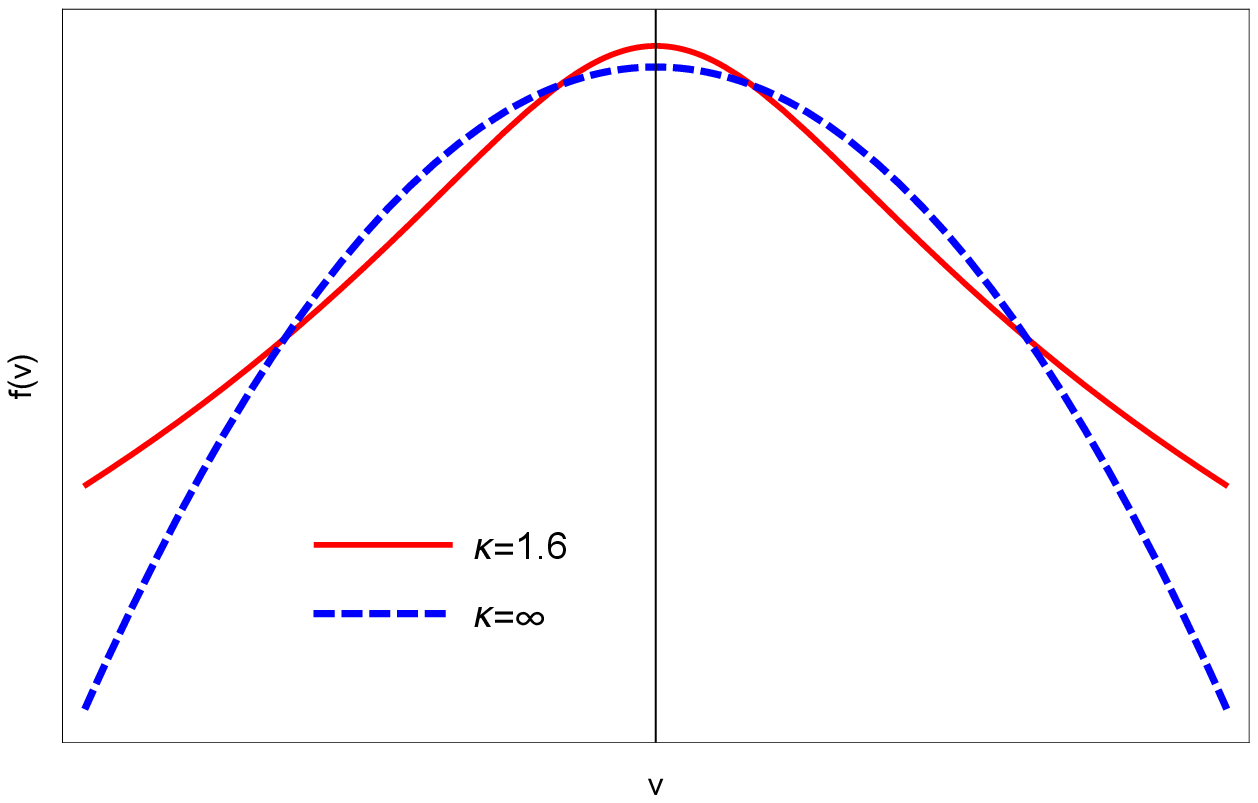}
	\caption{The schematic diagram of the comparison between kappa distribution and Maxwellian distribution in one dimension velocity distribution.}
	\label{kappa-vs-Maxwellian}
\end{figure}

In Fig.\ref{ft-1c:a} we find that the relaxation rate $(\partial f_\alpha/\partial t)_c$ hardly varies when the speed is from $10^{-2}\si{cm/s}$ to $10^{7} \si{cm/s}$, while decreases significantly when in the range of $10^7\si{cm/s} - 10^9 \si{cm/s}$ in the case of $\kappa_e=1.6$, $2$, and $5$. It is indicated that the collisions make stronger influences on the distribution of low energy particles than that of high energy particles. 
The missing fragment of the curves in the neighborhood of $v=10^7 \si{cm/s}$ in Fig.\ref{ft-1c:a} is the positive part of $(\partial f_e/\partial t)_c$. Thus, Fig.\ref{ft-1c:b} is drawn to exhibit this fragment more clearly in a relative small range of speed.
In Fig.\ref{ft-1c} the negative relaxation rate, namely $(\partial f_\alpha/\partial t)_c<0$, implies that the distribution decreases as the increment of the time and vice verse. As we know, the kappa distribution comparing with the Maxwellian one has larger values in small-velocity and superthermal range, which is displayed in a schematic diagram Fig.\ref{kappa-vs-Maxwellian}. Therefore, when the relaxation rate of the kappa distribution is negative in small-velocity and superthermal range as shown in Figs.\ref{ft-1c} and \ref{rft-1c}, we may think that the kappa distribution evolves "towards" Maxwellian distribution at initial time $t=0$. 
In addition, the larger the kappa parameter is, the smaller the relaxation rate $(\partial f_\alpha/\partial t)_c$ becomes. When the kappa parameter approach the infinity, $(\partial f_\alpha/\partial t)_c$ vanishes since the distribution reduces to the Maxwellian.
The relative relaxation rate $(\partial f_\alpha/\partial t)_c/f_\alpha$ in Fig.\ref{rft-1c} displays the same behaviors as the absolute rate of change.

\subsection{Plasma with two components}
We focus on electron-ion plasma in this part, which is a kind of two-component plasma. The FP collision term of electrons for such plasmas is divided into two parts,
\begin{equation}
	\left( \pdv[]{f_e}{t} \right)_c = \left( \pdv[]{f_e}{t} \right)_c^{ee}+\left( \pdv[]{f_e}{t} \right)_c^{ei},
	\label{eq-ft-2c}
\end{equation}
where the collision term with the superscript $ee$ denotes the contribution due to the electron-electron collision,
\begin{equation}
	\begin{split}
		\left( \pdv[]{f_e}{t} \right)_c^{ee} = \frac{4\pi}{3} f_e \Gamma_{ee} B_{\kappa,e} (1+A_{\kappa,e}v^2)^{-\kappa_e-1} \left\{ -\frac{3}{\kappa_e} -2(\kappa_e+1)A_{\kappa,e}v^2 \right. \\
		\times \left.\left[ {_2 F_1}\left( 1,\frac{3}{2}-\kappa_e;\frac{5}{2},-A_{\kappa,e}v^2 \right) -\frac{\kappa_e+2}{\kappa_e}{_2 F_1}\left( 1,\frac{5}{2}-\kappa_e;\frac{5}{2},-A_{\kappa,e}v^2 \right)  \right] \right\},
	\end{split}
	\label{eq-ft-2cee}
\end{equation}
and the collision term with $ei$ denotes the contribution from electron-ion collision,
\begin{equation}
	\begin{split}
		\left( \pdv[]{f_e}{t} \right)_c^{ei} =& \frac{4\pi}{3} f_e (1+A_{\kappa,e}v^2)^{-1} \Gamma_{ei} B_{\kappa,i} \frac{m_e}{m_i} (1+A_{\kappa,i}v^2)^{-\kappa_i} \left\{ 3 \left( \frac{1+A_{\kappa,e}v^2}{1+A_{\kappa,i}v^2} -\frac{\kappa_e+1}{\kappa_i}\frac{\kappa_i-\frac{3}{2}}{\kappa_e-\frac{3}{2}}\frac{T_i}{T_e} \right) \right.\\
						      &-2(\kappa_e+1)A_{\kappa,e}v^2 \left[ {_2 F_1}\left( 1,\frac{3}{2}-\kappa_i;\frac{5}{2},-A_{\kappa,i}v^2 \right)-\frac{\kappa_e+2}{\kappa_i}\frac{\kappa_i-\frac{3}{2}}{\kappa_e-\frac{3}{2}}\frac{T_i}{T_e} \frac{1+A_{\kappa,i}v^2}{1+A_{\kappa,e}v^2} \right. \\
						      &\left. \left. \times {_2 F_1}\left( 1,\frac{5}{2}-\kappa_i;\frac{5}{2},-A_{\kappa,i}v^2 \right) \right] \right\},
	\end{split}
	\label{eq-ft-2cei}
\end{equation}
which characterizes the relaxation of electrons in the background of kappa-distributed ions.  
The FP collision term of ions can be derived by replacing the physical quantities of electrons with those of ions symmetrically as well as the properties of the collision term. Therefore, only the properties of electron collision term are studied in this part for various kappa in the following.
It is worth to notice that the electron-electron collision term is the collision term of single-component plasma which has been already investigated in the previous subsection, which means the electron-ion collision term is the key point to discuss in this subsection. 
\subsubsection{The limit $\kappa_e \rightarrow +\infty$ and $\kappa_i \rightarrow +\infty$}
After taking the limit of $\kappa_e \rightarrow +\infty$ and $\kappa_i \rightarrow +\infty$, one yields that the FP collision term approach zero if $T_e=T_i$ as expected,
\begin{equation}
	\lim_{\substack{
			\kappa_e \rightarrow +\infty \\
			\kappa_i \rightarrow +\infty
		}}
	\left( \pdv[]{f_e}{t} \right)_c =0,
\end{equation}
which means the Maxwellian-distributed plasma is in the equilibrium as is known to all.
\subsubsection{The limit $\kappa_i \rightarrow \frac{3}{2}$}
\begin{figure}[ht]
	\centering
	\includegraphics[width=0.6\textwidth]{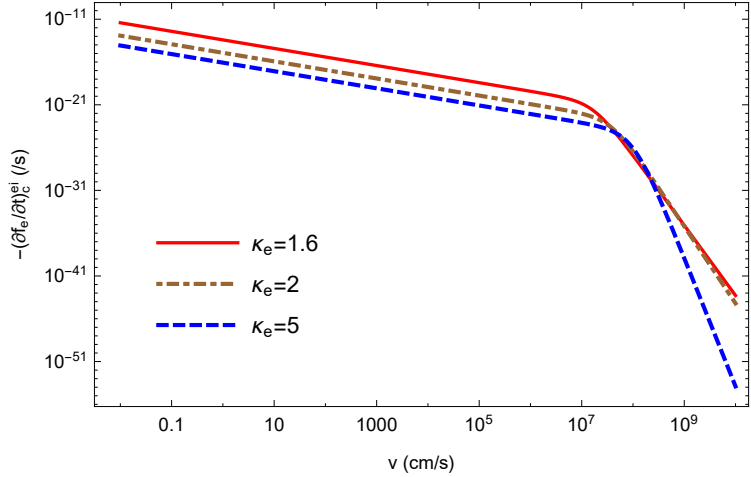}
	\caption{The absolute relaxation rate $(\partial f_e/\partial t)_c^{ei}$ due to the electron-ion collision in the background of kappa-distributed ions with $\kappa_i=3/2$.}
	\label{ft-2c32}
\end{figure}

\begin{figure}[ht]
	\centering
	\includegraphics[width=0.6\textwidth]{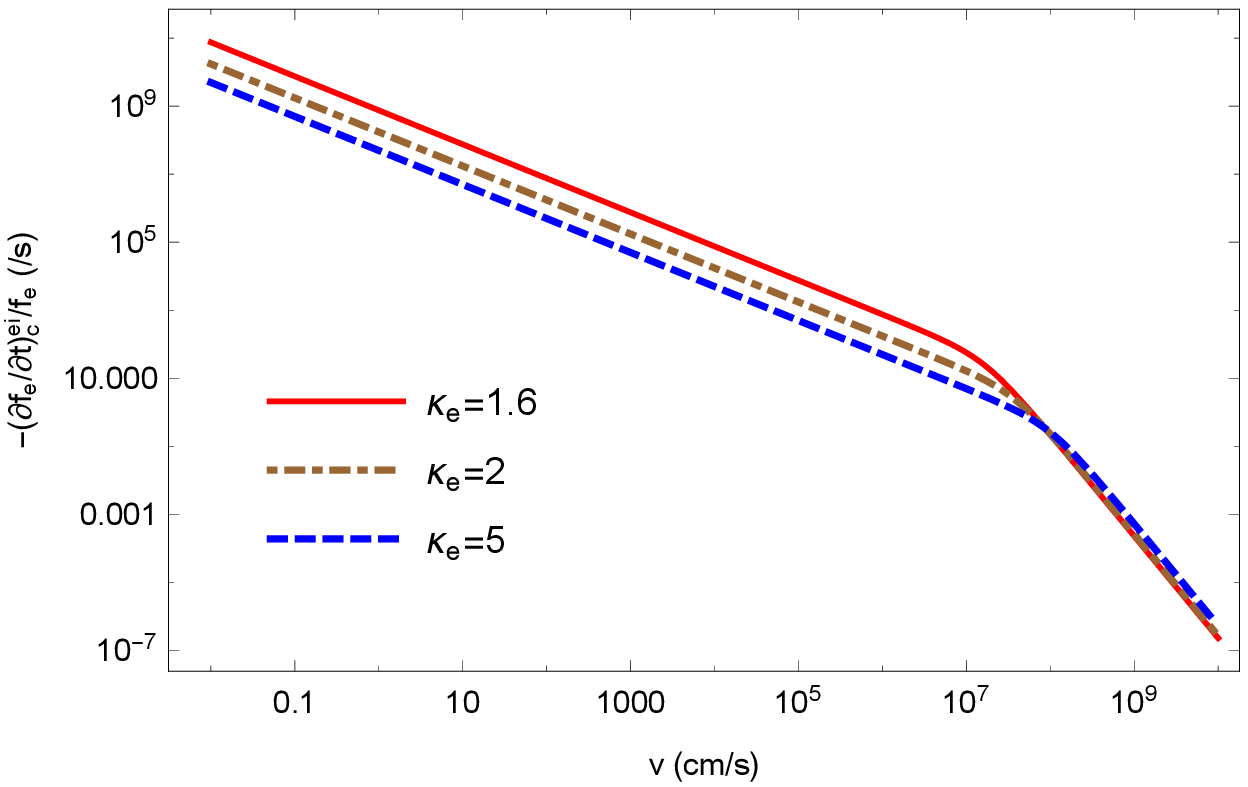}
	\caption{The relative relaxation rate $(\partial f_e/\partial t)_c/f_e$ due to the electron-ion collision in the background of kappa-distributed ions with $\kappa_i=3/2$.}
	\label{rft-2c32}
\end{figure}

In this case, the electron-ion collision term turns to be
\begin{equation}
	\lim_{\kappa_i \rightarrow \frac{3}{2}}\left( \pdv[]{f_e}{t} \right)_c^{ei} = -\Gamma_{ei}f_e{}\frac{2(\kappa_e+1)A_{\kappa,e}}{(1+A_{\kappa,e}v^2)v} \frac{m_e}{m_i},
	\label{eq-ft-2c32}
\end{equation}
which obviously cannot be zero for all $\vb v$. 
Therefore, Eq.\eqref{eq-ft-2c32} describes the relaxation of electrons in the background of kappa-distributed ions with $\kappa_i \rightarrow 3/2$. For the purpose of visualization we shall take the electron-proton plasma as an example for two-component plasma. The graph of the absolute and relative relaxation rate are drawn in Figs.\ref{ft-2c32} and \ref{rft-2c32}.  
The quantities of electron are set as same as those in the previous subsection; while for ions we also employ some typical values, namely the temperature $T_i = T_e/10 = 10^3 \si{K}$, the mass $m_i = 1.67\times10^{-24}\si{g}$, the charge of ions $q_i = q_e =4.8 \times 10^{-10}\si{statC}$ and the number density $n_i = n_e$. These figures show that the relaxation rate decreases when the speed increases significantly and becomes smaller for a larger $\kappa_e$-index. 
It is also illustrated that the relaxation rate is always negative for all range of speed $v$, which indicates the probability of all arbitrary $v$ reduces at initial time. 
\subsubsection{The limit $\kappa_i \rightarrow +\infty$}
\begin{figure}[ht]
	\centering
	\subfigure[The positive part of absolute rate for $\kappa_e=1.6$, $2$, and $5$ respectively.]{
	\includegraphics[width=0.6\textwidth]{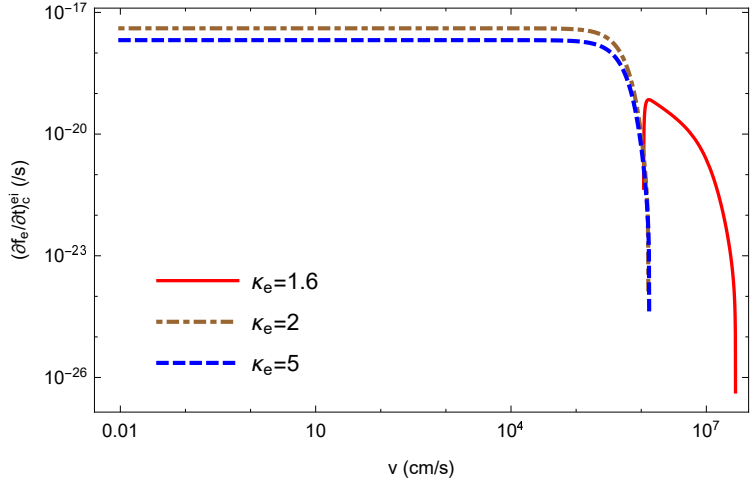}
	\label{ft-2cinf:a}
	}
	\subfigure[The negative part of absolute rate for $\kappa_e=1.6$, $2$, and $5$ respectively.]{
	\includegraphics[width=0.6\textwidth]{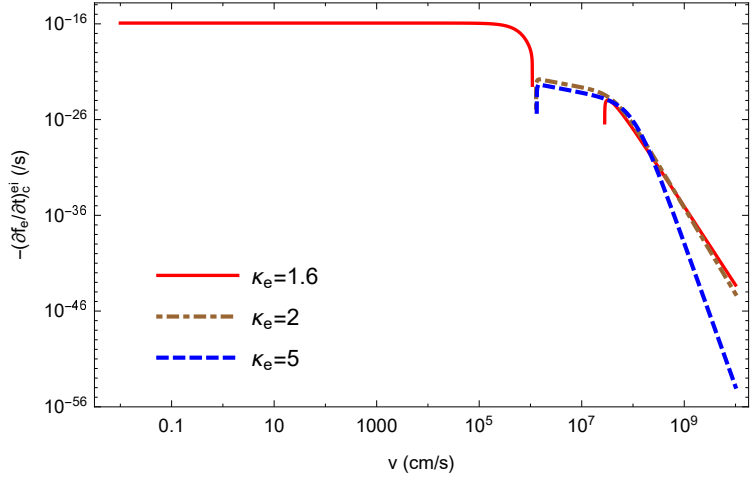}
	\label{ft-2cinf:b}
	}
	\caption{The absolute relaxation rate $(\partial f_e/\partial t)_c^{ei}$ due to the electron-ion collision in the background of Maxwellian-distributed ions.}
	\label{ft-2cinf}
\end{figure}

\begin{figure}[ht]
	\centering
	\subfigure[The positive part of relative rate for $\kappa_e=1.6$, $2$, and $5$ respectively.]{
	\includegraphics[width=0.6\textwidth]{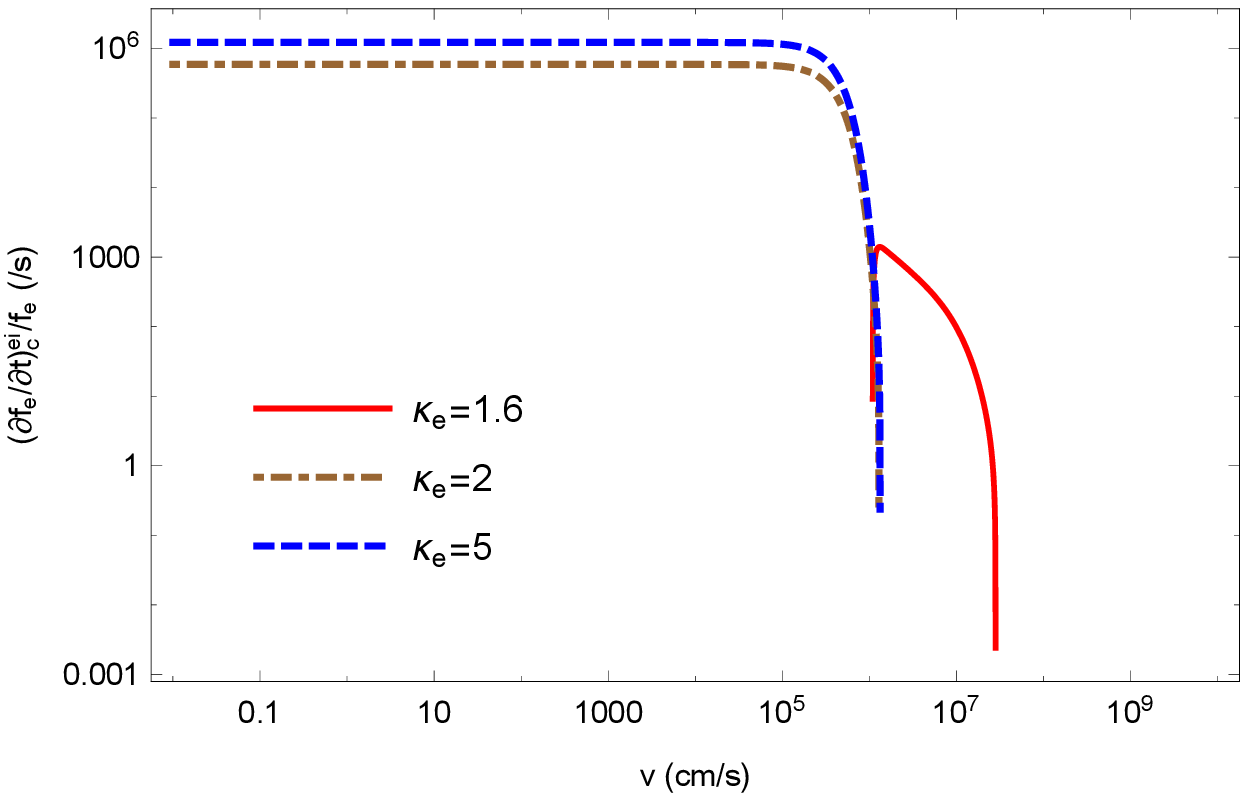}
	\label{rft-2cinf:a}
	}
	\subfigure[The negative part of relative rate for $\kappa_e=1.6$, $2$, and $5$ respectively.]{
	\includegraphics[width=0.6\textwidth]{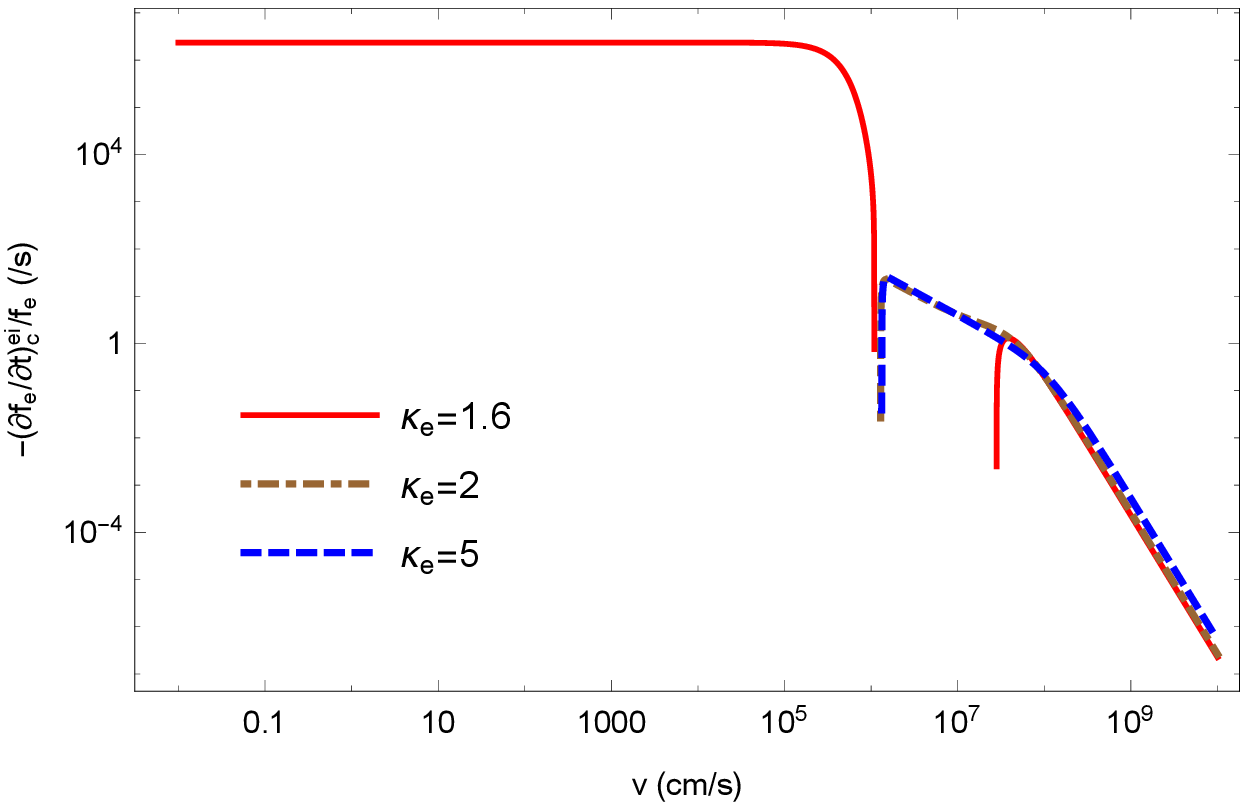}
	\label{rft-2cinf:b}
	}
	\caption{The relative relaxation rate $(\partial f_e/\partial t)_c^{ei}/f_e$ due to the electron-ion collision in the background of Maxwellian-distributed ions.}
	\label{rft-2cinf}
\end{figure}

When the ion distribution approaches the equilibrium, the FP collision term becomes,
\begin{equation}
	\begin{split}
		\lim_{\kappa_i \rightarrow +\infty}\left( \pdv[]{f_e}{t} \right)_c^{ei} = 4\pi \Gamma_{ei} \frac{f_e}{1+A_{\kappa,e}v^2}  \left( \frac{m_i}{2\pi k T_i} \right)^\frac{3}{2} e^{-\frac{m_i v^2}{2kT_i}} \frac{m_e}{m_i} \left\{ \left( 1+A_{\kappa,e}v^2-\frac{\kappa_e+1}{\kappa_e-\frac{3}{2}} \frac{T_i}{T_e} \right) \right. \\
	\left.	+(\kappa_e+1)A_{\kappa,e} \left[ \frac{2kT_i}{m_i}-\left( \frac{2kT_i}{m_i} \right)^\frac{3}{2}\frac{\sqrt{\pi} \Phi \left( \sqrt{\frac{m_i}{2kT_i}} v \right) }{2v e^{-\frac{m_i v^2}{2kT_i}}} \right]\left( 1-\frac{\kappa_e+2}{\kappa_e-\frac{3}{2}} \frac{T_i}{T_e}\frac{1}{1+A_{\kappa,e}v^2} \right)  \right\}.
	\end{split}
	\label{eq-ft-2cinf}
\end{equation}
The collision term \eqref{eq-ft-2cinf} is displayed in Figs.\ref{ft-2cinf} and \ref{rft-2cinf}. Both the absolute and relative relaxation rate are drawn in two subfigures in order to show the positive and negative parts in log-log scale. These graphs exhibit that the relaxation rate is significantly affected by the parameter $\kappa_e$. In the cases of $\kappa_e = 2$ and $\kappa_e =5$, the relaxation is positive in the range $v=10^{-2}\si{cm/s}-10^6\si{cm/s}$ and negative in the range $v=10^6\si{cm/s}-10^{10}\si{cm/s}$, which reveals that the probability of electrons $f_\alpha$ with small velocity increases during the time evolution while decreases for superthermal particles. It indicates that the electrons is decelerated due to the collision with Maxwellian-distributed ions and the shape of velocity distribution function would become more "narrow". However, in the case of $\kappa_e =1.6$, the numerical calculations give a different result that the relaxation rate is negative in the order of magnitude $v=10^{-2}\si{cm/s}-10^6\si{cm/s}$ and  $v=10^{8}\si{cm/s}-10^{10}\si{cm/s}$ and positive in the middle range $v=10^6\si{cm/s}-10^{10}\si{cm/s}$, which suggests the evolution "towards" the Maxwellian distribution. Therefore, the electron distributions with different $\kappa_e$-indexes can evolve towards different directions due to the electron-ion collision with Maxwellian-distributed ions. 

\subsubsection{The other values of $\kappa_i$}
\begin{figure}[ht]
	\centering
	\subfigure[The positive part of absolute relaxation rate in the range $v=10^{-2}\si{cm/s}-10^{6}\si{cm/s}$ for $\kappa_e=2$.]{
		\includegraphics[width=0.6\textwidth]{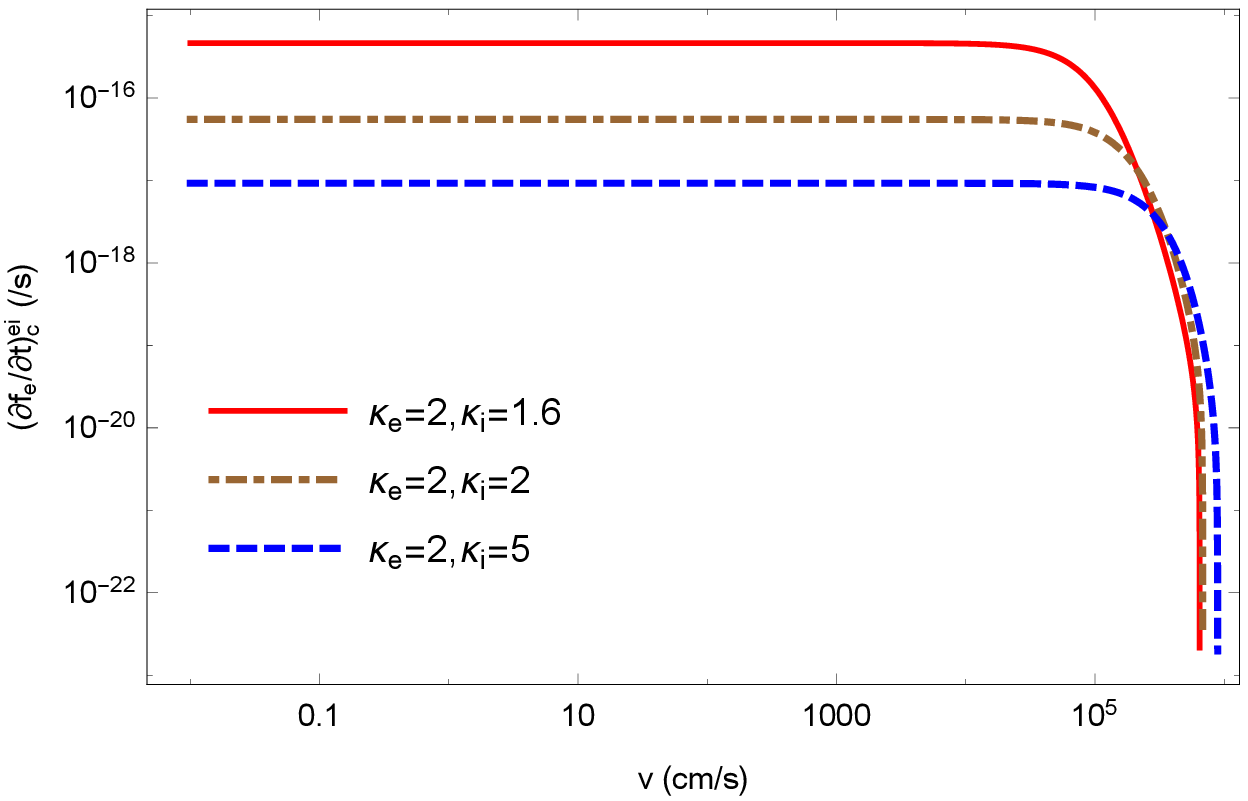}
		\label{ft-2cav:a}
	}
	\subfigure[The negative part of absolute relaxation rate in the range $v=10^{6}\si{cm/s}-10^{10}\si{cm/s}$ for $\kappa_e=2$.]{
		\includegraphics[width=0.6\textwidth]{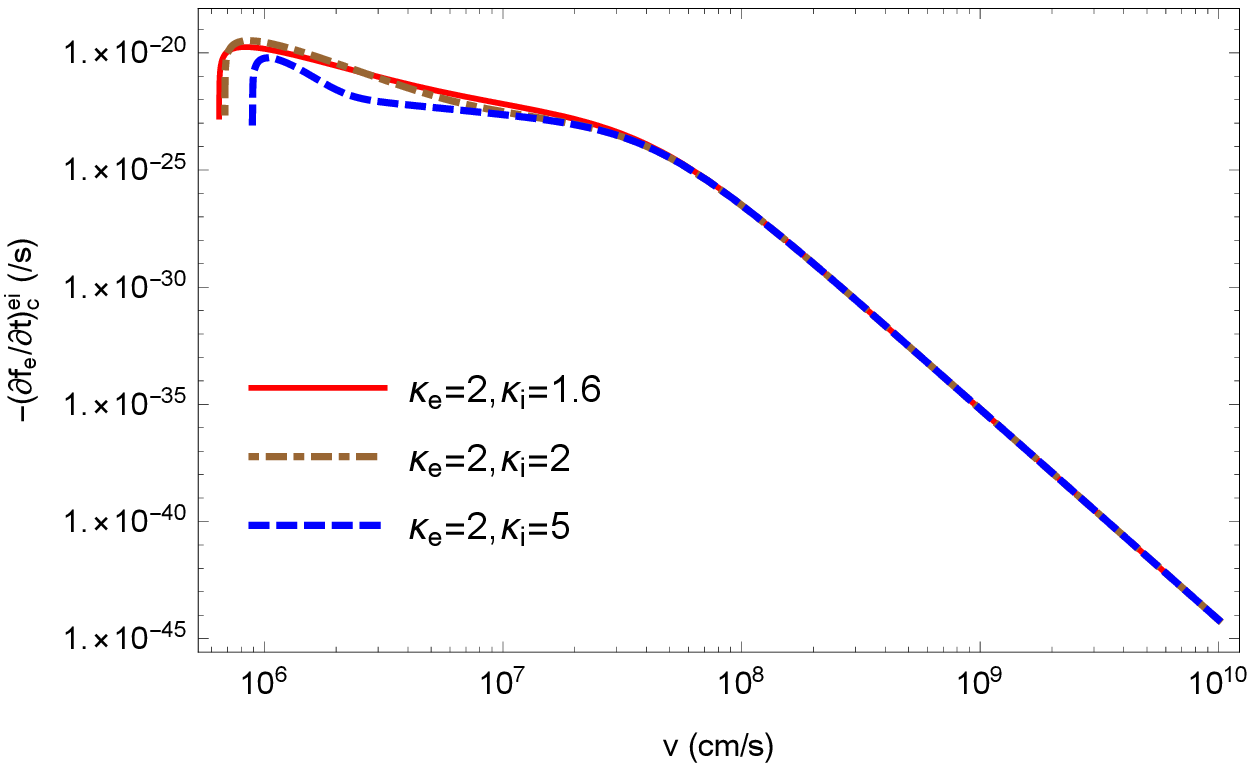}
		\label{ft-2cav:b}
	}
	\caption{The absolute relaxation rate $(\partial f_e/\partial t)_c^{ei}$ due to the electron-ion collision in the background of kappa-distributed ions with $\kappa_i=1.6$, $2$, and $5$ respectively.}
	\label{ft-2cav}
\end{figure}

\begin{figure}[ht]
	\centering
	\subfigure[The positive part of relative relaxation rate in the range $v=10^{-2}\si{cm/s}-10^{6}\si{cm/s}$ for $\kappa_e=2$.]{
	\includegraphics[width=0.7\textwidth]{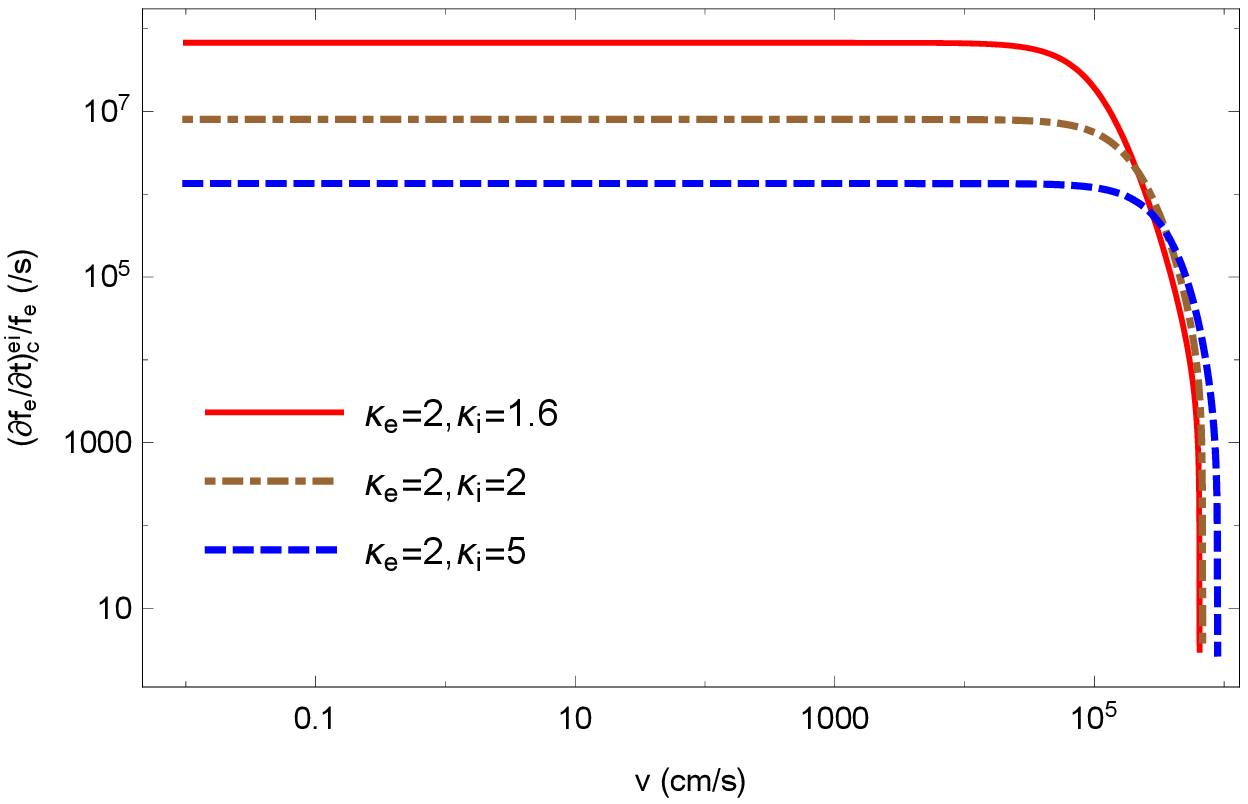}
	\label{rft-2cav:a}
	}
	\subfigure[The negative part of relative relaxation rate in the range $v=10^{6}\si{cm/s}-10^{10}\si{cm/s}$ for $\kappa_e=2$.]{
	\includegraphics[width=0.7\textwidth]{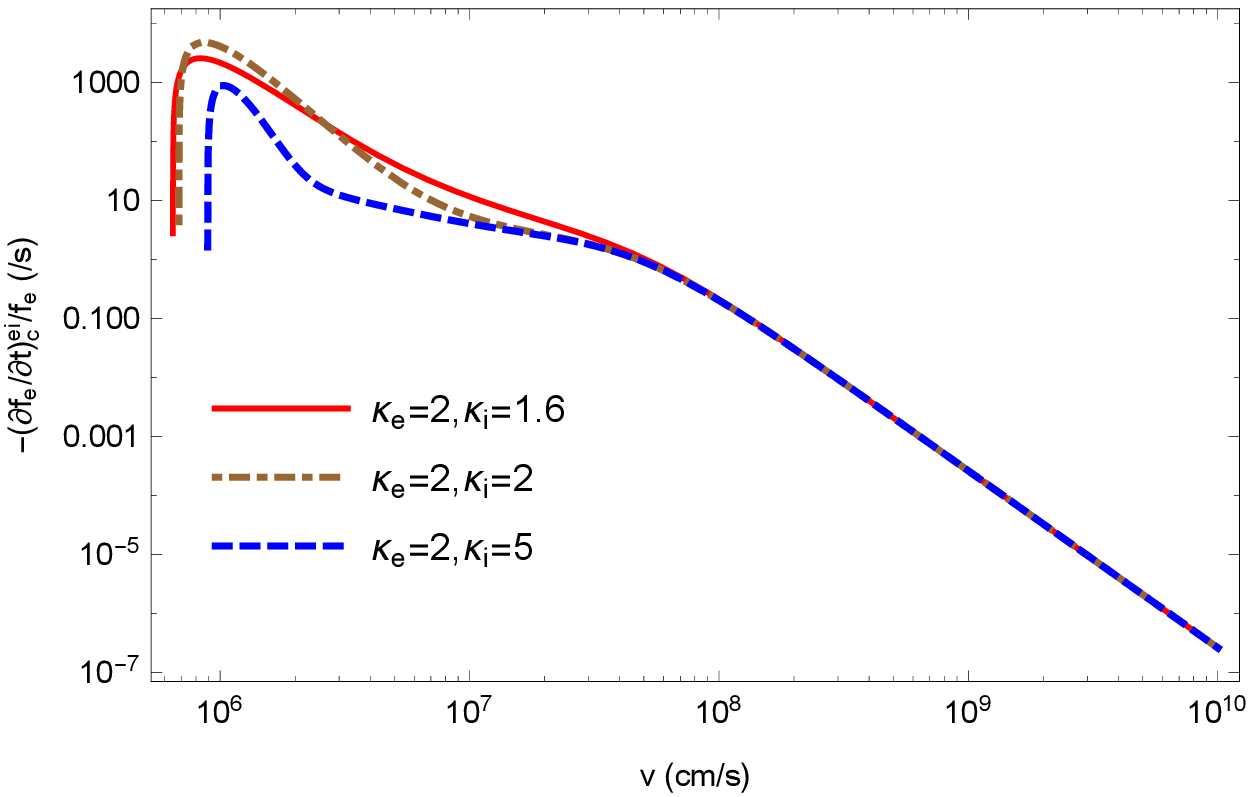}
	\label{rft-2cav:b}
	}
	\caption{The relative relaxation rate $(\partial f_e/\partial t)_c^{ei}/f_e$ due to the electron-ion collision in the background of kappa-distributed ions with $\kappa_i=1.6$, $2$, and $5$ respectively.}
	\label{rft-2cav}
\end{figure}

In this part, we numerically give the results \eqref{eq-ft-2cei} in Figs.\ref{ft-2cav} and \ref{rft-2cav}. The absolute and relative relaxation rate of electron kappa distribution are both drawn in the background of ion distributed with different kappa values. The kappa index of electron is set as $\kappa_e = 2$; while for the background ions the values of kappa are $\kappa_i = 1.6$, $2$ and $5$ respectively. In Figs.\ref{ft-2cav} and \ref{rft-2cav}, we find that the relaxation rate is positive in the range of $v = 10^{-2} \si{cm/s} - 10^6 \si{cm/s}$ and negative in the range of $v=10^6\si{cm/s}-10^{10}\si{cm/s}$, which means the electrons with high speed are decelerated by colliding with kappa-distributed ions. And additionally the smaller $\kappa_i$-index of the background ions is, the stronger decelerated effects turn to be.   %In those graphs, the relaxation rate behaviors similar properties in the case of $\kappa_i \rightarrow \infty$. The electron distribution also evolves backwards the Maxwellian distribution. %Combining the results in single-component plasma, we can conclude that the kappa-distributed electron evolves "towards" the Maxwellian distribution due to the collision with itself, while "backwards" the Maxwellian distribution due to the collision with ions.

\section{Conclusions}
In this work, we study fully ionized collisional plasma with kappa distribution by employing the FP equation with Rosenbluth potential. The results show that the kappa distribution cannot be a stationary unless the kappa parameter tends to infinity. Then we give the general analytical expression of the relaxation rate of kappa distribution \eqref{eq-fpc}. In order to analyze the relaxation rate more clearly, we discuss it in two special cases, i.e.\ , single-component plasma and electron-proton plasma. 

In single-component plasma, the relaxation rate is studied for $\kappa= +\infty$, $3/2$ and other values respectively. The results are derived analytically in Eqs. \eqref{eq-ft-1cinf}, \eqref{eq-ft-1c32}, and \eqref{eq-ft-1c}, and also shown in Figs.\ref{ft-1c} and \ref{rft-1c} numerically. The characterization of the relaxation rate in single-component is caused by the collisions with the single-component itself.

In electron-proton plasma, we focus on the relaxation rate contributed from the electron-ion collision, because the contribution from electron-electron collision has been studied in single-component case. Therefore, we investigate the electron relaxation rate in the background of kappa-distributed ions with $\kappa_i = +\infty$, $3/2$ and other values respectively. The results in different cases are derived in Eqs.\eqref{eq-ft-2cinf}, \eqref{eq-ft-2c32} and\eqref{eq-ft-2c}-\eqref{eq-ft-2cei}, and numerically shown in Figs.\ref{ft-2c32}-\ref{rft-2cav}. 

From these results, we can make the conclusions of the direction of the relaxation at initial time $t=0$. First, the kappa-distributed electron evolves "towards" the Maxwellian distribution due to the collision with itself. Second, the direction of the electron relaxation due to the collision with ions are complicated. When colliding with kappa-distributed ions, the electrons are decelerated.
When colliding with Maxwellian-distributed ions, kappa-distributed election with small enough $\kappa_e$-index evolves "towards" Maxwellian distribution such as the case of $\kappa_e=1.6$ while with other values of $\kappa_e$ the electrons are decelerated as exhibited in Fig.\ref{ft-2cinf} and \ref{rft-2cinf}. 
All these results illustrate that the kappa parameter significantly effects the relaxation rate due to the collision. 

\section{Acknowledgements}
This work was supported by the National Natural Science Foundation of China under grant No.11775156 and by the Special Foundation for Theoretical Physics under grant No.11747060 in the National Natural Science Foundation of China.
\begin{appendix}
	\section{Calculation of $g_{\alpha \beta}$ }
	\label{sec-app-g_ab}
	Taking the kappa distribution \eqref{eq-kappa-pdfa} into the definition of $g_{\alpha \beta} (\vb v)$ \eqref{eq-rp-g}, one gets
	\begin{equation}
		g_{\alpha \beta}(\vb v) = B_{\kappa,\beta} \int \left( 1+A_{\kappa,\beta} v'^2 \right)^{-\kappa_\beta-1} \left| \vb v' - \vb v \right| \dd \vb{v'}.
	\end{equation}
	Let $\vb w = \vb v' - \vb v$, and then the above equation turns into
	\begin{equation}
		g_{\alpha \beta}(\vb v)  = B_{\kappa,\beta} \int \left[ 1+A_{\kappa,\beta} (\vb w + \vb v)^2 \right]^{-\kappa_\beta-1}  w \dd \vb w.
		\label{int-g}
	\end{equation}
	We set z-axis in the direction of vector $\vb v$, and calculate the integral \eqref{int-g} in the spherical coordinate,
	\begin{equation}
		\begin{split}
			g_{\alpha \beta}(\vb v)  = 2\pi B_{\kappa,\beta} \int_0^\infty \dd w \left\{ w^3 \left[ 1+A_{\kappa,\beta} (w^2 + v^2) \right]^{-\kappa_\beta-1} \right\} \\
			\times \int_0^{2\pi} \dd \theta \left\{ \sin{\theta} \left[ 1+\frac{2A_{\kappa,\alpha}wv\cos{\theta}}{1+A_{\kappa,\alpha}(w^2+v^2)} \right]^{-\kappa_\beta-1} \right\},
		\end{split}
	\end{equation}
	where $\theta$ is the angle between $\vb w$ and $\vb v$. After working out the integral with respect to $\theta$, one has
	\begin{equation}
		g_{\alpha \beta}(\vb v)  = \frac{\pi B_{\kappa,\beta}}{\kappa_\beta v A_{\kappa,\beta}} \int_0^\infty  r^2 \left\{ [1+A_{\kappa,\alpha}(w-v)^2]^{-\kappa_\beta} - [1+A_{\kappa,\alpha}(w+v)^2]^{-\kappa_\beta} \right\} \dd r,
	\end{equation}
	which can be rearranged as
	\begin{equation}
		\begin{split}
			g_{\alpha \beta}(\vb{v}) =& \frac{\pi B_{\kappa,\beta}}{\kappa_\beta A_{\kappa,\beta}} \left[ \frac{2}{3} v^2 {_2 F_1}\left(\frac{3}{2},\kappa_\beta;\frac{5}{2},-A_{\kappa,\beta} v^2 \right) + 2 v^2 {_2 F_1}\left(\frac{1}{2},\kappa_\beta;\frac{3}{2},-A_{\kappa,\beta}v^2 \right) \right. \\
						  & \left. + \frac{2}{(\kappa_\beta-1) A_{\kappa,\beta}} (1+A_{\kappa,\beta} v^2)^{-\kappa_\beta+1} \right].
		\end{split}
	\end{equation}
	by employing the representation of hypergeometric function \eqref{eq-2f1}.

\end{appendix}
\bibliography{library}

\begin{thebibliography}{51}%
\makeatletter
\providecommand \@ifxundefined [1]{%
 \@ifx{#1\undefined}
}%
\providecommand \@ifnum [1]{%
 \ifnum #1\expandafter \@firstoftwo
 \else \expandafter \@secondoftwo
 \fi
}%
\providecommand \@ifx [1]{%
 \ifx #1\expandafter \@firstoftwo
 \else \expandafter \@secondoftwo
 \fi
}%
\providecommand \natexlab [1]{#1}%
\providecommand \enquote  [1]{``#1''}%
\providecommand \bibnamefont  [1]{#1}%
\providecommand \bibfnamefont [1]{#1}%
\providecommand \citenamefont [1]{#1}%
\providecommand \href@noop [0]{\@secondoftwo}%
\providecommand \href [0]{\begingroup \@sanitize@url \@href}%
\providecommand \@href[1]{\@@startlink{#1}\@@href}%
\providecommand \@@href[1]{\endgroup#1\@@endlink}%
\providecommand \@sanitize@url [0]{\catcode `\\12\catcode `\$12\catcode
  `\&12\catcode `\#12\catcode `\^12\catcode `\_12\catcode `\%12\relax}%
\providecommand \@@startlink[1]{}%
\providecommand \@@endlink[0]{}%
\providecommand \url  [0]{\begingroup\@sanitize@url \@url }%
\providecommand \@url [1]{\endgroup\@href {#1}{\urlprefix }}%
\providecommand \urlprefix  [0]{URL }%
\providecommand \Eprint [0]{\href }%
\providecommand \doibase [0]{http://dx.doi.org/}%
\providecommand \selectlanguage [0]{\@gobble}%
\providecommand \bibinfo  [0]{\@secondoftwo}%
\providecommand \bibfield  [0]{\@secondoftwo}%
\providecommand \translation [1]{[#1]}%
\providecommand \BibitemOpen [0]{}%
\providecommand \bibitemStop [0]{}%
\providecommand \bibitemNoStop [0]{.\EOS\space}%
\providecommand \EOS [0]{\spacefactor3000\relax}%
\providecommand \BibitemShut  [1]{\csname bibitem#1\endcsname}%
\let\auto@bib@innerbib\@empty
%</preamble>
\bibitem [{\citenamefont {Vasyliunas}(1968)}]{Vasyliunas1968JGR}%
  \BibitemOpen
  \bibfield  {author} {\bibinfo {author} {\bibfnamefont {V.~M.}\ \bibnamefont
  {Vasyliunas}},\ }\bibfield  {title} {\enquote {\bibinfo {title} {A survey of
  low-energy electrons in the evening sector of the magnetosphere with ogo 1
  and ogo 3},}\ }\href {\doibase 10.1029/ja073i009p02839} {\bibfield  {journal}
  {\bibinfo  {journal} {J. Geophys. Res.}\ }\textbf {\bibinfo {volume} {73}},\
  \bibinfo {pages} {2839--2884} (\bibinfo {year} {1968})}\BibitemShut {NoStop}%
\bibitem [{\citenamefont {Vocks}, \citenamefont {Mann},\ and\ \citenamefont
  {Rausche}(2008)}]{Vocks2008AA}%
  \BibitemOpen
  \bibfield  {author} {\bibinfo {author} {\bibfnamefont {C.}~\bibnamefont
  {Vocks}}, \bibinfo {author} {\bibfnamefont {G.}~\bibnamefont {Mann}}, \ and\
  \bibinfo {author} {\bibfnamefont {G.}~\bibnamefont {Rausche}},\ }\bibfield
  {title} {\enquote {\bibinfo {title} {Formation of suprathermal electron
  distributions in the quiet solar corona},}\ }\href {\doibase
  10.1051/0004-6361:20078826} {\bibfield  {journal} {\bibinfo  {journal}
  {Astronomy {\&} Astrophysics}\ }\textbf {\bibinfo {volume} {480}},\ \bibinfo
  {pages} {527--536} (\bibinfo {year} {2008})}\BibitemShut {NoStop}%
\bibitem [{\citenamefont {Cranmer}(2014)}]{Cranmer2014AJL}%
  \BibitemOpen
  \bibfield  {author} {\bibinfo {author} {\bibfnamefont {S.~R.}\ \bibnamefont
  {Cranmer}},\ }\bibfield  {title} {\enquote {\bibinfo {title} {Suprathermal
  electrons in the solar corona: Can nonlocal transport explain heliospheric
  charge states?}}\ }\href {\doibase 10.1088/2041-8205/791/2/L31} {\bibfield
  {journal} {\bibinfo  {journal} {Astrophys. J. Lett.}\ }\textbf {\bibinfo
  {volume} {791}},\ \bibinfo {pages} {L31} (\bibinfo {year} {2014})},\ \Eprint
  {http://arxiv.org/abs/1407.5941} {arXiv:1407.5941 [astro-ph.SR]} \BibitemShut
  {NoStop}%
\bibitem [{\citenamefont {Maksimovic}, \citenamefont {Pierrard},\ and\
  \citenamefont {Lemaire}(1997)}]{Maksimovic1997AA}%
  \BibitemOpen
  \bibfield  {author} {\bibinfo {author} {\bibfnamefont {M.}~\bibnamefont
  {Maksimovic}}, \bibinfo {author} {\bibfnamefont {V.}~\bibnamefont
  {Pierrard}}, \ and\ \bibinfo {author} {\bibfnamefont {J.~F.}\ \bibnamefont
  {Lemaire}},\ }\bibfield  {title} {\enquote {\bibinfo {title} {A kinetic model
  of the solar wind with kappa distribution functions in the corona},}\
  }\href@noop {} {\bibfield  {journal} {\bibinfo  {journal} {Astron.
  Astrophys}\ }\textbf {\bibinfo {volume} {324}},\ \bibinfo {pages} {725--734}
  (\bibinfo {year} {1997})}\BibitemShut {NoStop}%
\bibitem [{\citenamefont {Leubner}(2004{\natexlab{a}})}]{Leubner2004AJ}%
  \BibitemOpen
  \bibfield  {author} {\bibinfo {author} {\bibfnamefont {M.~P.}\ \bibnamefont
  {Leubner}},\ }\bibfield  {title} {\enquote {\bibinfo {title} {Core‐halo
  distribution functions: A natural equilibrium state in generalized
  thermostatistics},}\ }\href {\doibase 10.1086/381867} {\bibfield  {journal}
  {\bibinfo  {journal} {Astrophys. J.}\ }\textbf {\bibinfo {volume} {604}},\
  \bibinfo {pages} {469--478} (\bibinfo {year}
  {2004}{\natexlab{a}})}\BibitemShut {NoStop}%
\bibitem [{\citenamefont {Zouganelis}(2008)}]{Zouganelis2008JGRSP}%
  \BibitemOpen
  \bibfield  {author} {\bibinfo {author} {\bibfnamefont {I.}~\bibnamefont
  {Zouganelis}},\ }\bibfield  {title} {\enquote {\bibinfo {title} {Measuring
  suprathermal electron parameters in space plasmas: Implementation of the
  quasi-thermal noise spectroscopy with kappa distributions using in situ
  ulysses/{URAP} radio measurements in the solar wind},}\ }\href {\doibase
  10.1029/2007ja012979} {\bibfield  {journal} {\bibinfo  {journal} {J. Geophys.
  Res. Space Phys.}\ }\textbf {\bibinfo {volume} {113}},\ \bibinfo {pages}
  {A08111} (\bibinfo {year} {2008})}\BibitemShut {NoStop}%
\bibitem [{\citenamefont {{\v{S}}tver{\'{a}}k}\ \emph
  {et~al.}(2009)\citenamefont {{\v{S}}tver{\'{a}}k}, \citenamefont
  {Maksimovic}, \citenamefont {Tr{\'{a}}vn{\'{\i}}{\v{c}}ek}, \citenamefont
  {Marsch}, \citenamefont {Fazakerley},\ and\ \citenamefont
  {Scime}}]{Stverak2009JGRSP}%
  \BibitemOpen
  \bibfield  {author} {\bibinfo {author} {\bibfnamefont {{\v{S}}.}~\bibnamefont
  {{\v{S}}tver{\'{a}}k}}, \bibinfo {author} {\bibfnamefont {M.}~\bibnamefont
  {Maksimovic}}, \bibinfo {author} {\bibfnamefont {P.~M.}\ \bibnamefont
  {Tr{\'{a}}vn{\'{\i}}{\v{c}}ek}}, \bibinfo {author} {\bibfnamefont
  {E.}~\bibnamefont {Marsch}}, \bibinfo {author} {\bibfnamefont {A.~N.}\
  \bibnamefont {Fazakerley}}, \ and\ \bibinfo {author} {\bibfnamefont {E.~E.}\
  \bibnamefont {Scime}},\ }\bibfield  {title} {\enquote {\bibinfo {title}
  {Radial evolution of nonthermal electron populations in the low-latitude
  solar wind: Helios, cluster, and ulysses observations},}\ }\href {\doibase
  10.1029/2008ja013883} {\bibfield  {journal} {\bibinfo  {journal} {J. Geophys.
  Res. Space Phys.}\ }\textbf {\bibinfo {volume} {114}},\ \bibinfo {pages}
  {A05104} (\bibinfo {year} {2009})}\BibitemShut {NoStop}%
\bibitem [{\citenamefont {Livadiotis}\ \emph {et~al.}(2011)\citenamefont
  {Livadiotis}, \citenamefont {McComas}, \citenamefont {Dayeh}, \citenamefont
  {Funsten},\ and\ \citenamefont {Schwadron}}]{Livadiotis2011AJa}%
  \BibitemOpen
  \bibfield  {author} {\bibinfo {author} {\bibfnamefont {G.}~\bibnamefont
  {Livadiotis}}, \bibinfo {author} {\bibfnamefont {D.~J.}\ \bibnamefont
  {McComas}}, \bibinfo {author} {\bibfnamefont {M.~A.}\ \bibnamefont {Dayeh}},
  \bibinfo {author} {\bibfnamefont {H.~O.}\ \bibnamefont {Funsten}}, \ and\
  \bibinfo {author} {\bibfnamefont {N.~A.}\ \bibnamefont {Schwadron}},\
  }\bibfield  {title} {\enquote {\bibinfo {title} {{FIRST} {SKY} {MAP} {OF}
  {THE} {INNER} {HELIOSHEATH} {TEMPERATURE} {USINGIBEXSPECTRA}},}\ }\href
  {\doibase 10.1088/0004-637x/734/1/1} {\bibfield  {journal} {\bibinfo
  {journal} {Astrophys. J.}\ }\textbf {\bibinfo {volume} {734}},\ \bibinfo
  {pages} {1} (\bibinfo {year} {2011})}\BibitemShut {NoStop}%
\bibitem [{\citenamefont {Mauk}(2004)}]{Mauk2004JGR}%
  \BibitemOpen
  \bibfield  {author} {\bibinfo {author} {\bibfnamefont {B.~H.}\ \bibnamefont
  {Mauk}},\ }\bibfield  {title} {\enquote {\bibinfo {title} {Energetic ion
  characteristics and neutral gas interactions in jupiters magnetosphere},}\
  }\href {\doibase 10.1029/2003ja010270} {\bibfield  {journal} {\bibinfo
  {journal} {J. Geophys. Res.}\ }\textbf {\bibinfo {volume} {109}},\ \bibinfo
  {pages} {A09S12} (\bibinfo {year} {2004})}\BibitemShut {NoStop}%
\bibitem [{\citenamefont {Schippers}\ \emph {et~al.}(2008)\citenamefont
  {Schippers}, \citenamefont {Blanc}, \citenamefont {André}, \citenamefont
  {Dandouras}, \citenamefont {Lewis}, \citenamefont {Gilbert}, \citenamefont
  {Persoon}, \citenamefont {Krupp}, \citenamefont {Gurnett}, \citenamefont
  {Coates}, \citenamefont {Krimigis}, \citenamefont {Young},\ and\
  \citenamefont {Dougherty}}]{Schippers2008JGRSP}%
  \BibitemOpen
  \bibfield  {author} {\bibinfo {author} {\bibfnamefont {P.}~\bibnamefont
  {Schippers}}, \bibinfo {author} {\bibfnamefont {M.}~\bibnamefont {Blanc}},
  \bibinfo {author} {\bibfnamefont {N.}~\bibnamefont {André}}, \bibinfo
  {author} {\bibfnamefont {I.}~\bibnamefont {Dandouras}}, \bibinfo {author}
  {\bibfnamefont {G.~R.}\ \bibnamefont {Lewis}}, \bibinfo {author}
  {\bibfnamefont {L.~K.}\ \bibnamefont {Gilbert}}, \bibinfo {author}
  {\bibfnamefont {A.~M.}\ \bibnamefont {Persoon}}, \bibinfo {author}
  {\bibfnamefont {N.}~\bibnamefont {Krupp}}, \bibinfo {author} {\bibfnamefont
  {D.~A.}\ \bibnamefont {Gurnett}}, \bibinfo {author} {\bibfnamefont {A.~J.}\
  \bibnamefont {Coates}}, \bibinfo {author} {\bibfnamefont {S.~M.}\
  \bibnamefont {Krimigis}}, \bibinfo {author} {\bibfnamefont {D.~T.}\
  \bibnamefont {Young}}, \ and\ \bibinfo {author} {\bibfnamefont {M.~K.}\
  \bibnamefont {Dougherty}},\ }\bibfield  {title} {\enquote {\bibinfo {title}
  {Multi-instrument analysis of electron populations in saturns
  magnetosphere},}\ }\href {\doibase 10.1029/2008ja013098} {\bibfield
  {journal} {\bibinfo  {journal} {J. Geophys. Res. Space Phys.}\ }\textbf
  {\bibinfo {volume} {113}},\ \bibinfo {pages} {A07208} (\bibinfo {year}
  {2008})}\BibitemShut {NoStop}%
\bibitem [{\citenamefont {Dialynas}\ \emph {et~al.}(2009)\citenamefont
  {Dialynas}, \citenamefont {Krimigis}, \citenamefont {Mitchell}, \citenamefont
  {Hamilton}, \citenamefont {Krupp},\ and\ \citenamefont
  {Brandt}}]{Dialynas2009JGRSP}%
  \BibitemOpen
  \bibfield  {author} {\bibinfo {author} {\bibfnamefont {K.}~\bibnamefont
  {Dialynas}}, \bibinfo {author} {\bibfnamefont {S.~M.}\ \bibnamefont
  {Krimigis}}, \bibinfo {author} {\bibfnamefont {D.~G.}\ \bibnamefont
  {Mitchell}}, \bibinfo {author} {\bibfnamefont {D.~C.}\ \bibnamefont
  {Hamilton}}, \bibinfo {author} {\bibfnamefont {N.}~\bibnamefont {Krupp}}, \
  and\ \bibinfo {author} {\bibfnamefont {P.~C.}\ \bibnamefont {Brandt}},\
  }\bibfield  {title} {\enquote {\bibinfo {title} {Energetic ion spectral
  characteristics in the saturnian magnetosphere using cassini/{MIMI}
  measurements},}\ }\href {\doibase 10.1029/2008ja013761} {\bibfield  {journal}
  {\bibinfo  {journal} {J. Geophys. Res. Space Phys.}\ }\textbf {\bibinfo
  {volume} {114}},\ \bibinfo {pages} {A01212} (\bibinfo {year}
  {2009})}\BibitemShut {NoStop}%
\bibitem [{\citenamefont {Gougam}\ and\ \citenamefont
  {Tribeche}(2011)}]{Gougam2011PP}%
  \BibitemOpen
  \bibfield  {author} {\bibinfo {author} {\bibfnamefont {L.~A.}\ \bibnamefont
  {Gougam}}\ and\ \bibinfo {author} {\bibfnamefont {M.}~\bibnamefont
  {Tribeche}},\ }\bibfield  {title} {\enquote {\bibinfo {title} {Debye
  shielding in a nonextensive plasma},}\ }\href {\doibase 10.1063/1.3577599}
  {\bibfield  {journal} {\bibinfo  {journal} {Phys. Plasmas}\ }\textbf
  {\bibinfo {volume} {18}},\ \bibinfo {pages} {062102} (\bibinfo {year}
  {2011})}\BibitemShut {NoStop}%
\bibitem [{\citenamefont {Livadiotis}\ and\ \citenamefont
  {McComas}(2014)}]{Livadiotis2014JPP}%
  \BibitemOpen
  \bibfield  {author} {\bibinfo {author} {\bibfnamefont {G.}~\bibnamefont
  {Livadiotis}}\ and\ \bibinfo {author} {\bibfnamefont {D.~J.}\ \bibnamefont
  {McComas}},\ }\bibfield  {title} {\enquote {\bibinfo {title} {Electrostatic
  shielding in plasmas and the physical meaning of the debye length},}\ }\href
  {\doibase 10.1017/S0022377813001335} {\bibfield  {journal} {\bibinfo
  {journal} {J. Plasma Phys.}\ }\textbf {\bibinfo {volume} {80}},\ \bibinfo
  {pages} {341--378} (\bibinfo {year} {2014})}\BibitemShut {NoStop}%
\bibitem [{\citenamefont {Livadiotis}\ and\ \citenamefont
  {McComas}(2009)}]{Livadiotis2009JGRSP}%
  \BibitemOpen
  \bibfield  {author} {\bibinfo {author} {\bibfnamefont {G.}~\bibnamefont
  {Livadiotis}}\ and\ \bibinfo {author} {\bibfnamefont {D.~J.}\ \bibnamefont
  {McComas}},\ }\bibfield  {title} {\enquote {\bibinfo {title} {Beyond kappa
  distributions: Exploiting tsallis statistical mechanics in space plasmas},}\
  }\href {\doibase 10.1029/2009JA014352} {\bibfield  {journal} {\bibinfo
  {journal} {J. Geophys. Res. Space Phys.}\ }\textbf {\bibinfo {volume}
  {114}},\ \bibinfo {pages} {A11105} (\bibinfo {year} {2009})}\BibitemShut
  {NoStop}%
\bibitem [{\citenamefont {Livadiotis}(2014)}]{Livadiotis2014E}%
  \BibitemOpen
  \bibfield  {author} {\bibinfo {author} {\bibfnamefont {G.}~\bibnamefont
  {Livadiotis}},\ }\bibfield  {title} {\enquote {\bibinfo {title} {Lagrangian
  temperature: Derivation and physical meaning for systems described by kappa
  distributions},}\ }\href {\doibase 10.3390/e16084290} {\bibfield  {journal}
  {\bibinfo  {journal} {Entropy}\ }\textbf {\bibinfo {volume} {16}},\ \bibinfo
  {pages} {4290--4308} (\bibinfo {year} {2014})}\BibitemShut {NoStop}%
\bibitem [{\citenamefont {Amour}\ and\ \citenamefont
  {Tribeche}(2010)}]{Amour2010PP}%
  \BibitemOpen
  \bibfield  {author} {\bibinfo {author} {\bibfnamefont {R.}~\bibnamefont
  {Amour}}\ and\ \bibinfo {author} {\bibfnamefont {M.}~\bibnamefont
  {Tribeche}},\ }\bibfield  {title} {\enquote {\bibinfo {title} {Variable
  charge dust acoustic solitary waves in a dusty plasma with a q-nonextensive
  electron velocity distribution},}\ }\href {\doibase 10.1063/1.3428538}
  {\bibfield  {journal} {\bibinfo  {journal} {Phys. Plasmas}\ }\textbf
  {\bibinfo {volume} {17}},\ \bibinfo {pages} {063702} (\bibinfo {year}
  {2010})}\BibitemShut {NoStop}%
\bibitem [{\citenamefont {Tribeche}, \citenamefont {Amour},\ and\ \citenamefont
  {Shukla}(2012)}]{Tribeche2012PRE}%
  \BibitemOpen
  \bibfield  {author} {\bibinfo {author} {\bibfnamefont {M.}~\bibnamefont
  {Tribeche}}, \bibinfo {author} {\bibfnamefont {R.}~\bibnamefont {Amour}}, \
  and\ \bibinfo {author} {\bibfnamefont {P.~K.}\ \bibnamefont {Shukla}},\
  }\bibfield  {title} {\enquote {\bibinfo {title} {Ion acoustic solitary waves
  in a plasma with nonthermal electrons featuring tsallis distribution},}\
  }\href {\doibase 10.1103/physreve.85.037401} {\bibfield  {journal} {\bibinfo
  {journal} {Phys. Rev. E}\ }\textbf {\bibinfo {volume} {85}},\ \bibinfo
  {pages} {037401} (\bibinfo {year} {2012})}\BibitemShut {NoStop}%
\bibitem [{\citenamefont {Das}\ and\ \citenamefont
  {Karmakar}(2018)}]{Das2018AA}%
  \BibitemOpen
  \bibfield  {author} {\bibinfo {author} {\bibfnamefont {P.}~\bibnamefont
  {Das}}\ and\ \bibinfo {author} {\bibfnamefont {P.~K.}\ \bibnamefont
  {Karmakar}},\ }\bibfield  {title} {\enquote {\bibinfo {title} {Nonlinear
  waves in viscoelastic magnetized complex astroplasmas with polarized
  dust-charge variations},}\ }\href {\doibase 10.1063/1.5011386} {\bibfield
  {journal} {\bibinfo  {journal} {AIP Adv.}\ }\textbf {\bibinfo {volume} {8}},\
  \bibinfo {pages} {015010} (\bibinfo {year} {2018})}\BibitemShut {NoStop}%
\bibitem [{\citenamefont {Liu}, \citenamefont {Liu},\ and\ \citenamefont
  {Du}(2009)}]{Liu2009PP}%
  \BibitemOpen
  \bibfield  {author} {\bibinfo {author} {\bibfnamefont {Z.}~\bibnamefont
  {Liu}}, \bibinfo {author} {\bibfnamefont {L.}~\bibnamefont {Liu}}, \ and\
  \bibinfo {author} {\bibfnamefont {J.}~\bibnamefont {Du}},\ }\bibfield
  {title} {\enquote {\bibinfo {title} {A nonextensive approach for the
  instability of current-driven ion-acoustic waves in space plasmas},}\ }\href
  {\doibase 10.1063/1.3176516} {\bibfield  {journal} {\bibinfo  {journal}
  {Phys. Plasmas}\ }\textbf {\bibinfo {volume} {16}},\ \bibinfo {pages}
  {072111} (\bibinfo {year} {2009})},\ \Eprint {http://arxiv.org/abs/0907.1966}
  {arXiv:0907.1966} \BibitemShut {NoStop}%
\bibitem [{\citenamefont {Liu}\ and\ \citenamefont {Du}(2009)}]{Liu2009PPa}%
  \BibitemOpen
  \bibfield  {author} {\bibinfo {author} {\bibfnamefont {Z.}~\bibnamefont
  {Liu}}\ and\ \bibinfo {author} {\bibfnamefont {J.}~\bibnamefont {Du}},\
  }\bibfield  {title} {\enquote {\bibinfo {title} {Dust acoustic instability
  driven by drifting ions and electrons in the dust plasma with lorentzian
  kappa distribution},}\ }\href {\doibase 10.1063/1.3274459} {\bibfield
  {journal} {\bibinfo  {journal} {Phys. Plasmas}\ }\textbf {\bibinfo {volume}
  {16}},\ \bibinfo {pages} {123707} (\bibinfo {year} {2009})}\BibitemShut
  {NoStop}%
\bibitem [{\citenamefont {Mace}\ and\ \citenamefont
  {Sydora}(2010)}]{Mace2010JGRSP}%
  \BibitemOpen
  \bibfield  {author} {\bibinfo {author} {\bibfnamefont {R.~L.}\ \bibnamefont
  {Mace}}\ and\ \bibinfo {author} {\bibfnamefont {R.~D.}\ \bibnamefont
  {Sydora}},\ }\bibfield  {title} {\enquote {\bibinfo {title} {Parallel
  whistler instability in a plasma with an anisotropic bi-kappa
  distribution},}\ }\href {\doibase 10.1029/2009ja015064} {\bibfield  {journal}
  {\bibinfo  {journal} {J. Geophys. Res. Space Phys.}\ }\textbf {\bibinfo
  {volume} {115}},\ \bibinfo {pages} {A07206} (\bibinfo {year}
  {2010})}\BibitemShut {NoStop}%
\bibitem [{\citenamefont {Eslami}, \citenamefont {Mottaghizadeh},\ and\
  \citenamefont {Pakzad}(2011)}]{Eslami2011PP}%
  \BibitemOpen
  \bibfield  {author} {\bibinfo {author} {\bibfnamefont {P.}~\bibnamefont
  {Eslami}}, \bibinfo {author} {\bibfnamefont {M.}~\bibnamefont
  {Mottaghizadeh}}, \ and\ \bibinfo {author} {\bibfnamefont {H.~R.}\
  \bibnamefont {Pakzad}},\ }\bibfield  {title} {\enquote {\bibinfo {title}
  {Modulational instability of ion acoustic waves in e-p-i plasmas with
  electrons and positrons following a q-nonextensive distribution},}\ }\href
  {\doibase 10.1063/1.3646318} {\bibfield  {journal} {\bibinfo  {journal}
  {Phys. Plasmas}\ }\textbf {\bibinfo {volume} {18}},\ \bibinfo {pages}
  {102313} (\bibinfo {year} {2011})}\BibitemShut {NoStop}%
\bibitem [{\citenamefont {Lazar}, \citenamefont {Poedts},\ and\ \citenamefont
  {Schlickeiser}(2010)}]{Lazar2010MNRAS}%
  \BibitemOpen
  \bibfield  {author} {\bibinfo {author} {\bibfnamefont {M.}~\bibnamefont
  {Lazar}}, \bibinfo {author} {\bibfnamefont {S.}~\bibnamefont {Poedts}}, \
  and\ \bibinfo {author} {\bibfnamefont {R.}~\bibnamefont {Schlickeiser}},\
  }\bibfield  {title} {\enquote {\bibinfo {title} {Instability of the parallel
  electromagnetic modes in kappa distributed plasmas - i. electron
  whistler-cyclotron modes},}\ }\href {\doibase
  10.1111/j.1365-2966.2010.17472.x} {\bibfield  {journal} {\bibinfo  {journal}
  {Mon. Not. R. Astron. Soc.}\ }\textbf {\bibinfo {volume} {410}},\ \bibinfo
  {pages} {663--670} (\bibinfo {year} {2010})}\BibitemShut {NoStop}%
\bibitem [{\citenamefont {Lazar}\ and\ \citenamefont
  {Poedts}(2013)}]{Lazar2013MNRAS}%
  \BibitemOpen
  \bibfield  {author} {\bibinfo {author} {\bibfnamefont {M.}~\bibnamefont
  {Lazar}}\ and\ \bibinfo {author} {\bibfnamefont {S.}~\bibnamefont {Poedts}},\
  }\bibfield  {title} {\enquote {\bibinfo {title} {Instability of the parallel
  electromagnetic modes in kappa distributed plasmas {\textendash} {II}.
  electromagnetic ion{\textendash}cyclotron modes},}\ }\href {\doibase
  10.1093/mnras/stt1914} {\bibfield  {journal} {\bibinfo  {journal} {Mon. Not.
  R. Astron. Soc.}\ }\textbf {\bibinfo {volume} {437}},\ \bibinfo {pages}
  {641--648} (\bibinfo {year} {2013})}\BibitemShut {NoStop}%
\bibitem [{\citenamefont {Vi{\~{n}}as}\ \emph {et~al.}(2015)\citenamefont
  {Vi{\~{n}}as}, \citenamefont {Moya}, \citenamefont {Navarro}, \citenamefont
  {Valdivia}, \citenamefont {Araneda},\ and\ \citenamefont
  {Mu{\~{n}}oz}}]{Vinas2015JGRSP}%
  \BibitemOpen
  \bibfield  {author} {\bibinfo {author} {\bibfnamefont {A.~F.}\ \bibnamefont
  {Vi{\~{n}}as}}, \bibinfo {author} {\bibfnamefont {P.~S.}\ \bibnamefont
  {Moya}}, \bibinfo {author} {\bibfnamefont {R.~E.}\ \bibnamefont {Navarro}},
  \bibinfo {author} {\bibfnamefont {J.~A.}\ \bibnamefont {Valdivia}}, \bibinfo
  {author} {\bibfnamefont {J.~A.}\ \bibnamefont {Araneda}}, \ and\ \bibinfo
  {author} {\bibfnamefont {V.}~\bibnamefont {Mu{\~{n}}oz}},\ }\bibfield
  {title} {\enquote {\bibinfo {title} {Electromagnetic fluctuations of the
  whistler-cyclotron and firehose instabilities in a maxwellian and
  tsallis-kappa-like plasma},}\ }\href {\doibase 10.1002/2014ja020554}
  {\bibfield  {journal} {\bibinfo  {journal} {J. Geophys. Res. Space Phys.}\
  }\textbf {\bibinfo {volume} {120}},\ \bibinfo {pages} {3307--3317} (\bibinfo
  {year} {2015})}\BibitemShut {NoStop}%
\bibitem [{\citenamefont {Baluku}\ and\ \citenamefont
  {Hellberg}(2015)}]{Baluku2015PP}%
  \BibitemOpen
  \bibfield  {author} {\bibinfo {author} {\bibfnamefont {T.~K.}\ \bibnamefont
  {Baluku}}\ and\ \bibinfo {author} {\bibfnamefont {M.~A.}\ \bibnamefont
  {Hellberg}},\ }\bibfield  {title} {\enquote {\bibinfo {title} {Kinetic theory
  of dust ion acoustic waves in a kappa-distributed plasma},}\ }\href {\doibase
  10.1063/1.4927581} {\bibfield  {journal} {\bibinfo  {journal} {Phys.
  Plasmas}\ }\textbf {\bibinfo {volume} {22}},\ \bibinfo {pages} {083701}
  (\bibinfo {year} {2015})}\BibitemShut {NoStop}%
\bibitem [{\citenamefont {Du}(2013)}]{Du2013PP}%
  \BibitemOpen
  \bibfield  {author} {\bibinfo {author} {\bibfnamefont {J.}~\bibnamefont
  {Du}},\ }\bibfield  {title} {\enquote {\bibinfo {title} {Transport
  coefficients in lorentz plasmas with the power-law kappa-distribution},}\
  }\href {\doibase 10.1063/1.4820799} {\bibfield  {journal} {\bibinfo
  {journal} {Phys. Plasmas}\ }\textbf {\bibinfo {volume} {20}},\ \bibinfo
  {pages} {092901} (\bibinfo {year} {2013})},\ \Eprint
  {http://arxiv.org/abs/1307.3849} {arXiv:1307.3849} \BibitemShut {NoStop}%
\bibitem [{\citenamefont {Abbasi}\ and\ \citenamefont
  {Esfandyari-Kalejahi}(2016)}]{Abbasi2016PP}%
  \BibitemOpen
  \bibfield  {author} {\bibinfo {author} {\bibfnamefont {Z.~E.}\ \bibnamefont
  {Abbasi}}\ and\ \bibinfo {author} {\bibfnamefont {A.}~\bibnamefont
  {Esfandyari-Kalejahi}},\ }\bibfield  {title} {\enquote {\bibinfo {title} {The
  effect of non-extensive particles on slowing down and diffusion of a
  particles beam in the plasma},}\ }\href {\doibase 10.1063/1.4956451}
  {\bibfield  {journal} {\bibinfo  {journal} {Phys. Plasmas}\ }\textbf
  {\bibinfo {volume} {23}},\ \bibinfo {pages} {073112} (\bibinfo {year}
  {2016})}\BibitemShut {NoStop}%
\bibitem [{\citenamefont {Wang}\ and\ \citenamefont {Du}(2017)}]{Wang2017PP}%
  \BibitemOpen
  \bibfield  {author} {\bibinfo {author} {\bibfnamefont {L.}~\bibnamefont
  {Wang}}\ and\ \bibinfo {author} {\bibfnamefont {J.}~\bibnamefont {Du}},\
  }\bibfield  {title} {\enquote {\bibinfo {title} {The diffusion of charged
  particles in the weakly ionized plasma with power-law kappa-distributions},}\
  }\href {\doibase 10.1063/1.4996775} {\bibfield  {journal} {\bibinfo
  {journal} {Phys. Plasmas}\ }\textbf {\bibinfo {volume} {24}},\ \bibinfo
  {pages} {102305} (\bibinfo {year} {2017})}\BibitemShut {NoStop}%
\bibitem [{\citenamefont {Ebne~Abbasi}, \citenamefont {Esfandyari-Kalejahi},\
  and\ \citenamefont {Khaledi}(2017)}]{EbneAbbasi2017ASS}%
  \BibitemOpen
  \bibfield  {author} {\bibinfo {author} {\bibfnamefont {Z.}~\bibnamefont
  {Ebne~Abbasi}}, \bibinfo {author} {\bibfnamefont {A.}~\bibnamefont
  {Esfandyari-Kalejahi}}, \ and\ \bibinfo {author} {\bibfnamefont
  {P.}~\bibnamefont {Khaledi}},\ }\bibfield  {title} {\enquote {\bibinfo
  {title} {The collision times and transport coefficients of a fully ionized
  plasma with superthermal particles},}\ }\href {\doibase
  10.1007/s10509-017-3081-4} {\bibfield  {journal} {\bibinfo  {journal}
  {Astrophys. Space Sci.}\ }\textbf {\bibinfo {volume} {362}},\ \bibinfo
  {pages} {103} (\bibinfo {year} {2017})}\BibitemShut {NoStop}%
\bibitem [{\citenamefont {Wang}\ and\ \citenamefont {Du}(2018)}]{Wang2018PP}%
  \BibitemOpen
  \bibfield  {author} {\bibinfo {author} {\bibfnamefont {Y.}~\bibnamefont
  {Wang}}\ and\ \bibinfo {author} {\bibfnamefont {J.}~\bibnamefont {Du}},\
  }\bibfield  {title} {\enquote {\bibinfo {title} {The viscosity of charged
  particles in the weakly ionized plasma with power-law distributions},}\
  }\href {\doibase 10.1063/1.5023030} {\bibfield  {journal} {\bibinfo
  {journal} {Phys. Plasmas}\ }\textbf {\bibinfo {volume} {25}},\ \bibinfo
  {pages} {062309} (\bibinfo {year} {2018})}\BibitemShut {NoStop}%
\bibitem [{\citenamefont {Marsch}\ and\ \citenamefont
  {Livi}(1985)}]{Marsch1985PF}%
  \BibitemOpen
  \bibfield  {author} {\bibinfo {author} {\bibfnamefont {E.}~\bibnamefont
  {Marsch}}\ and\ \bibinfo {author} {\bibfnamefont {S.}~\bibnamefont {Livi}},\
  }\bibfield  {title} {\enquote {\bibinfo {title} {Coulomb collision rates for
  self-similar and kappa distributions},}\ }\href {\doibase 10.1063/1.864971}
  {\bibfield  {journal} {\bibinfo  {journal} {Phys. Fluids}\ }\textbf {\bibinfo
  {volume} {28}},\ \bibinfo {pages} {1379} (\bibinfo {year}
  {1985})}\BibitemShut {NoStop}%
\bibitem [{\citenamefont {Hau}\ and\ \citenamefont {Fu}(2007)}]{Hau2007PP}%
  \BibitemOpen
  \bibfield  {author} {\bibinfo {author} {\bibfnamefont {L.-N.}\ \bibnamefont
  {Hau}}\ and\ \bibinfo {author} {\bibfnamefont {W.-Z.}\ \bibnamefont {Fu}},\
  }\bibfield  {title} {\enquote {\bibinfo {title} {Mathematical and physical
  aspects of kappa velocity distribution},}\ }\href {\doibase
  10.1063/1.2779283} {\bibfield  {journal} {\bibinfo  {journal} {Phys.
  Plasmas}\ }\textbf {\bibinfo {volume} {14}},\ \bibinfo {pages} {110702}
  (\bibinfo {year} {2007})}\BibitemShut {NoStop}%
\bibitem [{\citenamefont {Gong}\ and\ \citenamefont
  {Du}(2012{\natexlab{a}})}]{Gong2012PP}%
  \BibitemOpen
  \bibfield  {author} {\bibinfo {author} {\bibfnamefont {J.}~\bibnamefont
  {Gong}}\ and\ \bibinfo {author} {\bibfnamefont {J.}~\bibnamefont {Du}},\
  }\bibfield  {title} {\enquote {\bibinfo {title} {Secondary electron emissions
  and dust charging currents in the nonequilibrium dusty plasma with power-law
  distributions},}\ }\href {\doibase 10.1063/1.4729684} {\bibfield  {journal}
  {\bibinfo  {journal} {Phys. Plasmas}\ }\textbf {\bibinfo {volume} {19}},\
  \bibinfo {pages} {063703} (\bibinfo {year} {2012}{\natexlab{a}})},\ \Eprint
  {http://arxiv.org/abs/1202.0636} {arXiv:1202.0636} \BibitemShut {NoStop}%
\bibitem [{\citenamefont {Gong}\ and\ \citenamefont
  {Du}(2012{\natexlab{b}})}]{Gong2012PPa}%
  \BibitemOpen
  \bibfield  {author} {\bibinfo {author} {\bibfnamefont {J.}~\bibnamefont
  {Gong}}\ and\ \bibinfo {author} {\bibfnamefont {J.}~\bibnamefont {Du}},\
  }\bibfield  {title} {\enquote {\bibinfo {title} {Dust charging processes in
  the nonequilibrium dusty plasma with nonextensive power-law distribution},}\
  }\href {\doibase 10.1063/1.3682051} {\bibfield  {journal} {\bibinfo
  {journal} {Phys. Plasmas}\ }\textbf {\bibinfo {volume} {19}},\ \bibinfo
  {pages} {023704} (\bibinfo {year} {2012}{\natexlab{b}})},\ \Eprint
  {http://arxiv.org/abs/1202.0636} {arXiv:1202.0636} \BibitemShut {NoStop}%
\bibitem [{\citenamefont {Cranmer}(1998)}]{Cranmer1998AJ}%
  \BibitemOpen
  \bibfield  {author} {\bibinfo {author} {\bibfnamefont {S.~R.}\ \bibnamefont
  {Cranmer}},\ }\bibfield  {title} {\enquote {\bibinfo {title} {Non-maxwellian
  redistribution in solar coronal ly emission},}\ }\href {\doibase
  10.1086/306415} {\bibfield  {journal} {\bibinfo  {journal} {Astrophys. J.}\
  }\textbf {\bibinfo {volume} {508}},\ \bibinfo {pages} {925--939} (\bibinfo
  {year} {1998})}\BibitemShut {NoStop}%
\bibitem [{\citenamefont {Leubner}(2002)}]{Leubner2002ASS}%
  \BibitemOpen
  \bibfield  {author} {\bibinfo {author} {\bibfnamefont {M.~P.}\ \bibnamefont
  {Leubner}},\ }\bibfield  {title} {\enquote {\bibinfo {title} {A nonextensive
  entropy approach to kappa-distributions},}\ }\href {\doibase
  10.1023/A:1020990413487} {\bibfield  {journal} {\bibinfo  {journal}
  {Astrophys. Space Sci.}\ }\textbf {\bibinfo {volume} {282}},\ \bibinfo
  {pages} {573--579} (\bibinfo {year} {2002})},\ \Eprint
  {http://arxiv.org/abs/0111444} {arXiv:0111444 [astro-ph]} \BibitemShut
  {NoStop}%
\bibitem [{\citenamefont {Leubner}(2004{\natexlab{b}})}]{Leubner2004PP}%
  \BibitemOpen
  \bibfield  {author} {\bibinfo {author} {\bibfnamefont {M.~P.}\ \bibnamefont
  {Leubner}},\ }\bibfield  {title} {\enquote {\bibinfo {title} {Fundamental
  issues on kappa-distributions in space plasmas and interplanetary proton
  distributions},}\ }\href {\doibase 10.1063/1.1667501} {\bibfield  {journal}
  {\bibinfo  {journal} {Phys. Plasmas}\ }\textbf {\bibinfo {volume} {11}},\
  \bibinfo {pages} {1308--1316} (\bibinfo {year}
  {2004}{\natexlab{b}})}\BibitemShut {NoStop}%
\bibitem [{\citenamefont {Livadiotis}(2015)}]{Livadiotis2015JGRSP}%
  \BibitemOpen
  \bibfield  {author} {\bibinfo {author} {\bibfnamefont {G.}~\bibnamefont
  {Livadiotis}},\ }\bibfield  {title} {\enquote {\bibinfo {title} {Introduction
  to special section on origins and properties of kappa distributions:
  Statistical background and properties of kappa distributions in space
  plasmas},}\ }\href {\doibase 10.1002/2014ja020825} {\bibfield  {journal}
  {\bibinfo  {journal} {J. Geophys. Res. Space Phys.}\ }\textbf {\bibinfo
  {volume} {120}},\ \bibinfo {pages} {1607--1619} (\bibinfo {year}
  {2015})}\BibitemShut {NoStop}%
\bibitem [{\citenamefont {Du}(2004)}]{Du2004PLA}%
  \BibitemOpen
  \bibfield  {author} {\bibinfo {author} {\bibfnamefont {J.}~\bibnamefont
  {Du}},\ }\bibfield  {title} {\enquote {\bibinfo {title} {Nonextensivity in
  nonequilibrium plasma systems with coulombian long-range interactions},}\
  }\href {\doibase 10.1016/j.physleta.2004.07.010} {\bibfield  {journal}
  {\bibinfo  {journal} {Phys. Lett. A}\ }\textbf {\bibinfo {volume} {329}},\
  \bibinfo {pages} {262--267} (\bibinfo {year} {2004})},\ \Eprint
  {http://arxiv.org/abs/0404602} {arXiv:0404602 [cond-mat]} \BibitemShut
  {NoStop}%
\bibitem [{\citenamefont {Yu}\ and\ \citenamefont {Du}(2014)}]{Yu2014AP}%
  \BibitemOpen
  \bibfield  {author} {\bibinfo {author} {\bibfnamefont {H.}~\bibnamefont
  {Yu}}\ and\ \bibinfo {author} {\bibfnamefont {J.}~\bibnamefont {Du}},\
  }\bibfield  {title} {\enquote {\bibinfo {title} {The nonextensive parameter
  for nonequilibrium electron gas in an electromagnetic field},}\ }\href
  {\doibase 10.1016/j.aop.2014.07.028} {\bibfield  {journal} {\bibinfo
  {journal} {Ann. Phys.}\ }\textbf {\bibinfo {volume} {350}},\ \bibinfo {pages}
  {302--309} (\bibinfo {year} {2014})}\BibitemShut {NoStop}%
\bibitem [{\citenamefont {Du}(2012)}]{Du2012JSMTE}%
  \BibitemOpen
  \bibfield  {author} {\bibinfo {author} {\bibfnamefont {J.-L.}\ \bibnamefont
  {Du}},\ }\bibfield  {title} {\enquote {\bibinfo {title} {Power-law
  distributions and fluctuation-dissipation relation in the stochastic dynamics
  of two-variable langevin equations},}\ }\href {\doibase
  10.1088/1742-5468/2012/02/P02006} {\bibfield  {journal} {\bibinfo  {journal}
  {Journal of Statistical Mechanics: Theory and Experiment}\ }\textbf {\bibinfo
  {volume} {2012}},\ \bibinfo {pages} {P02006} (\bibinfo {year} {2012})},\
  \Eprint {http://arxiv.org/abs/1202.0707} {1202.0707} \BibitemShut {NoStop}%
\bibitem [{\citenamefont {Guo}\ and\ \citenamefont {Du}(2013)}]{Guo2013JSMTE}%
  \BibitemOpen
  \bibfield  {author} {\bibinfo {author} {\bibfnamefont {R.}~\bibnamefont
  {Guo}}\ and\ \bibinfo {author} {\bibfnamefont {J.}~\bibnamefont {Du}},\
  }\bibfield  {title} {\enquote {\bibinfo {title} {Power-law behaviors from the
  two-variable langevin equation: Ito's and stratonovich's fokker–planck
  equations},}\ }\href {\doibase 10.1088/1742-5468/2013/02/P02015} {\bibfield
  {journal} {\bibinfo  {journal} {J. Stat. Mech: Theory Exp.}\ }\textbf
  {\bibinfo {volume} {2013}},\ \bibinfo {pages} {P02015} (\bibinfo {year}
  {2013})},\ \Eprint {http://arxiv.org/abs/1212.3980} {arXiv:1212.3980}
  \BibitemShut {NoStop}%
\bibitem [{\citenamefont {Guo}\ and\ \citenamefont {Du}(2014)}]{Guo2014PAa}%
  \BibitemOpen
  \bibfield  {author} {\bibinfo {author} {\bibfnamefont {R.}~\bibnamefont
  {Guo}}\ and\ \bibinfo {author} {\bibfnamefont {J.}~\bibnamefont {Du}},\
  }\bibfield  {title} {\enquote {\bibinfo {title} {Are power-law distributions
  an equilibrium distribution or a stationary nonequilibrium distribution?}}\
  }\href {\doibase 10.1016/j.physa.2014.03.056} {\bibfield  {journal} {\bibinfo
   {journal} {Physica A}\ }\textbf {\bibinfo {volume} {406}},\ \bibinfo {pages}
  {281--286} (\bibinfo {year} {2014})},\ \Eprint
  {http://arxiv.org/abs/1403.5441} {arXiv:1403.5441} \BibitemShut {NoStop}%
\bibitem [{\citenamefont {Hasegawa}, \citenamefont {Mima},\ and\ \citenamefont
  {Duong-van}(1985)}]{Hasegawa1985PRL}%
  \BibitemOpen
  \bibfield  {author} {\bibinfo {author} {\bibfnamefont {A.}~\bibnamefont
  {Hasegawa}}, \bibinfo {author} {\bibfnamefont {K.}~\bibnamefont {Mima}}, \
  and\ \bibinfo {author} {\bibfnamefont {M.}~\bibnamefont {Duong-van}},\
  }\bibfield  {title} {\enquote {\bibinfo {title} {Plasma distribution function
  in a superthermal radiation field},}\ }\href {\doibase
  10.1103/PhysRevLett.54.2608} {\bibfield  {journal} {\bibinfo  {journal}
  {Phys. Rev. Lett.}\ }\textbf {\bibinfo {volume} {54}},\ \bibinfo {pages}
  {2608--2610} (\bibinfo {year} {1985})}\BibitemShut {NoStop}%
\bibitem [{\citenamefont {Yoon}(2014)}]{Yoon2014JGRSP}%
  \BibitemOpen
  \bibfield  {author} {\bibinfo {author} {\bibfnamefont {P.~H.}\ \bibnamefont
  {Yoon}},\ }\bibfield  {title} {\enquote {\bibinfo {title} {Electron kappa
  distribution and quasi-thermal noise},}\ }\href {\doibase
  10.1002/2014ja020353} {\bibfield  {journal} {\bibinfo  {journal} {J. Geophys.
  Res. Space Phys.}\ }\textbf {\bibinfo {volume} {119}},\ \bibinfo {pages}
  {7074--7087} (\bibinfo {year} {2014})}\BibitemShut {NoStop}%
\bibitem [{\citenamefont {Bian}\ \emph {et~al.}(2014)\citenamefont {Bian},
  \citenamefont {Emslie}, \citenamefont {Stackhouse},\ and\ \citenamefont
  {Kontar}}]{Bian2014AJ}%
  \BibitemOpen
  \bibfield  {author} {\bibinfo {author} {\bibfnamefont {N.~H.}\ \bibnamefont
  {Bian}}, \bibinfo {author} {\bibfnamefont {A.~G.}\ \bibnamefont {Emslie}},
  \bibinfo {author} {\bibfnamefont {D.~J.}\ \bibnamefont {Stackhouse}}, \ and\
  \bibinfo {author} {\bibfnamefont {E.~P.}\ \bibnamefont {Kontar}},\ }\bibfield
   {title} {\enquote {\bibinfo {title} {{THE} {FORMATION} {OF}
  {KAPPA}-{DISTRIBUTION} {ACCELERATED} {ELECTRON} {POPULATIONS} {IN} {SOLAR}
  {FLARES}},}\ }\href {\doibase 10.1088/0004-637x/796/2/142} {\bibfield
  {journal} {\bibinfo  {journal} {Astrophys. J.}\ }\textbf {\bibinfo {volume}
  {796}},\ \bibinfo {pages} {142} (\bibinfo {year} {2014})}\BibitemShut
  {NoStop}%
\bibitem [{\citenamefont {Krall}\ and\ \citenamefont
  {Trivelpiece}(1973)}]{Krall1973}%
  \BibitemOpen
  \bibfield  {author} {\bibinfo {author} {\bibfnamefont {N.~A.}\ \bibnamefont
  {Krall}}\ and\ \bibinfo {author} {\bibfnamefont {A.~W.}\ \bibnamefont
  {Trivelpiece}},\ }\href
  {https://www.amazon.com/Principles-Plasma-Physics-Nicholas-Krall/dp/0070353468?SubscriptionId=AKIAIOBINVZYXZQZ2U3A&tag=chimbori05-20&linkCode=xm2&camp=2025&creative=165953&creativeASIN=0070353468}
  {\emph {\bibinfo {title} {Principles of Plasma Physics}}}\ (\bibinfo
  {publisher} {McGraw-Hill Inc.,US},\ \bibinfo {year} {1973})\BibitemShut
  {NoStop}%
\bibitem [{\citenamefont {Gurnett}\ and\ \citenamefont
  {Bhattacharjee}(2005)}]{Gurnett2005}%
  \BibitemOpen
  \bibfield  {author} {\bibinfo {author} {\bibfnamefont {D.~A.}\ \bibnamefont
  {Gurnett}}\ and\ \bibinfo {author} {\bibfnamefont {A.}~\bibnamefont
  {Bhattacharjee}},\ }\href
  {https://www.amazon.com/Introduction-Plasma-Physics-Laboratory-Applications/dp/0521364833?SubscriptionId=AKIAIOBINVZYXZQZ2U3A&tag=chimbori05-20&linkCode=xm2&camp=2025&creative=165953&creativeASIN=0521364833}
  {\emph {\bibinfo {title} {Introduction to Plasma Physics: With Space and
  Laboratory Applications}}}\ (\bibinfo  {publisher} {Cambridge University
  Press},\ \bibinfo {year} {2005})\BibitemShut {NoStop}%
\bibitem [{\citenamefont {Du}\ \emph {et~al.}(2018)\citenamefont {Du},
  \citenamefont {Guo}, \citenamefont {Liu},\ and\ \citenamefont
  {Du}}]{Du2018CPP}%
  \BibitemOpen
  \bibfield  {author} {\bibinfo {author} {\bibfnamefont {J.}~\bibnamefont
  {Du}}, \bibinfo {author} {\bibfnamefont {R.}~\bibnamefont {Guo}}, \bibinfo
  {author} {\bibfnamefont {Z.}~\bibnamefont {Liu}}, \ and\ \bibinfo {author}
  {\bibfnamefont {S.}~\bibnamefont {Du}},\ }\bibfield  {title} {\enquote
  {\bibinfo {title} {Slowing down of charged particles in dusty plasmas with
  power-law kappa-distributions},}\ }\href {\doibase 10.1002/ctpp.201800046}
  {\bibfield  {journal} {\bibinfo  {journal} {Contrib. Plasma Phys.}\ }
  (\bibinfo {year} {2018}),\ 10.1002/ctpp.201800046},\ \Eprint
  {http://arxiv.org/abs/1708.04525v1} {1708.04525v1} \BibitemShut {NoStop}%
\bibitem [{\citenamefont {Olver}\ \emph {et~al.}(2010)\citenamefont {Olver},
  \citenamefont {Lozier}, \citenamefont {Boisvert},\ and\ \citenamefont
  {Clark}}]{Olver2010NIST}%
  \BibitemOpen
  \bibfield  {author} {\bibinfo {author} {\bibfnamefont {F.~W.}\ \bibnamefont
  {Olver}}, \bibinfo {author} {\bibfnamefont {D.~W.}\ \bibnamefont {Lozier}},
  \bibinfo {author} {\bibfnamefont {R.~F.}\ \bibnamefont {Boisvert}}, \ and\
  \bibinfo {author} {\bibfnamefont {C.~W.}\ \bibnamefont {Clark}},\ }\href
  {www.cambridge.org/catalogue/catalogue.asp?isbn=9780521192255} {\emph
  {\bibinfo {title} {NIST Handbook of Mathematical Functions}}}\ (\bibinfo
  {publisher} {Cambridge University Press},\ \bibinfo {year}
  {2010})\BibitemShut {NoStop}%
\end{thebibliography}%
\end{document}